\documentclass[nonacm, sigconf]{acmart}

\usepackage{commath}
\usepackage[utf8]{inputenc}
\usepackage{graphicx}
\usepackage{subfig}
\usepackage[colorinlistoftodos]{todonotes}
\usepackage{comment}
\usepackage{breqn}
\usepackage{amsmath,amsfonts,amsthm,bm}
\usepackage{ulem}
\usepackage{multirow}
\usepackage{caption}
\usepackage{dsfont}
\graphicspath{{./images/}}

\theoremstyle{plain}

\AtBeginDocument{%
  \providecommand\BibTeX{{%
    \normalfont B\kern-0.5em{\scshape i\kern-0.25em b}\kern-0.8em\TeX}}}

\setcopyright{none}
\renewcommand\footnotetextcopyrightpermission[1]{}
\begin{document}
    
\title{Correcting for Selection Bias in Learning-to-rank Systems}

\author{Zohreh Ovaisi}
\email{zovais2@uic.edu}
\affiliation{%
\institution{University of Illinois at Chicago}
}

\author{Ragib Ahsan}
\email{rahsan@uic.edu}
\affiliation{%
\institution{University of Illinois at Chicago}
}

\author{Yifan Zhang}
\email{zhyifan3@mail3.sysu.edu.cn }
\affiliation{%
\institution{Sun Yat-sen University}
}

\author{Kathryn Vasilaky}
\email{kvasilak@calpoly.edu}
\affiliation{%
\institution{California Polytechnic State University}
}

\author{Elena Zheleva}
\email{ezheleva@uic.edu}
\affiliation{%
\institution{University of Illinois at Chicago}
}

\renewcommand{\shortauthors}{Zohreh Ovaisi, Ragib Ahsan, Yifan Zhang, Kathryn Vasilaky, Elena Zheleva}

\begin{abstract}
Click data collected by modern recommendation systems are an important source of observational data that can be utilized to train learning-to-rank (LTR) systems. However, these data suffer from a number of biases that can result in poor performance for LTR systems. Recent methods for bias correction in such systems mostly focus on position bias, the fact that higher ranked results (e.g., top search engine results) are more likely to be clicked even if they are not the most relevant results given a user’s query. Less attention has been paid to correcting for selection bias, which occurs because clicked documents are reflective of what documents have been shown to the user in the first place. Here, we propose new counterfactual approaches which adapt Heckman's two-stage method and accounts for selection and position bias in LTR systems. Our empirical evaluation shows that 
our proposed methods are much more robust to noise and have better accuracy compared to existing unbiased LTR algorithms, especially when there is moderate to no position bias. %

\end{abstract}

%
%
%
%
%
%
%
%
%
%
%
%
%
%
%
%
%
%
%
%
%
%
%
%

%
%
%
%



%
%
%

\maketitle


\section{Introduction}

The abundance of data found online has inspired new lines of inquiry about human behavior and the development of machine-learning algorithms that learn individual preferences from such data. Patterns in such data are often driven by the underlying algorithms supporting online platforms, rather than naturally-occurring user behavior. %
For example, interaction data from social media news feeds, such as user clicks and comments on posts, reflect not only latent user interests but also news feed personalization and what the underlying algorithms chose to show to users in the first place. 
Such data in turn are used to train new news feed algorithms, propagating the bias further~\cite{chaney-recsys18}. This can lead to phenomena such as filter bubbles and echo chambers and can challenge the validity of social science research that relies on found data~\cite{japec-poq15,lazer-science14}.

One of the places where these biases surface is in personalized recommender systems whose goal is to learn user preferences from available interaction data. These systems typically rely on learning procedures to estimate the parameters of \textit{new ranking algorithms} that are capable of ranking items based on inferred user preferences, in a process known as \textit{learning-to-rank (LTR)}~\cite{liu-springer11}. %
Much of the work on unbiasing the parameter estimation for learning-to-rank systems has focused on position bias~\cite{joachims-wsdm17}, the bias caused by the position where a result was displayed to a user. Position bias makes higher ranked results (e.g., top search engine results) more likely to be clicked even if they are not the most relevant. 

Algorithms that correct for position bias typically assume that all relevant results have non-zero probability of being observed (and thus clicked) by the user and focus on %
boosting the relevance of lower ranked relevant results~\cite{joachims-wsdm17}. However, users rarely have the chance to observe all relevant results, either because the system chose to show a truncated list of top $k$ recommended results or because users do not spend the time to peruse through tens to hundreds of ranked results. In this case, lower ranked, relevant results have zero probability of being observed (and clicked) and never get the chance to be boosted in LTR systems. This leads to selection bias in clicked results which is the focus of our work. 

Here, we frame the problem of learning to rank as a counterfactual problem of predicting whether a document would have been clicked had it been observed. In order to recover from selection bias for clicked documents, we focus on identifying the relevant documents that were never shown to users. Our formulation is different from
previous counterfactual formulations which correct for position bias and study the likelihood of a document being clicked had it been placed in a higher position %
given that it was placed in a lower position~\cite{joachims-wsdm17}. %

Here, we propose a general framework for recovering from selection bias that stems from both limited choices given to users and position bias. First, we propose $Heckman^{rank}$, an algorithm for addressing selection bias in the context of learning-to-rank systems. By adapting Heckman's two-stage method, an econometric tool for addressing selection bias, we account for the limited choice given to users and the fact that some items are more likely to be shown to a user than others. Because this correction method is very general, it is applicable to any type of selection bias in which the system's decision to show documents can be learned from features. Because $Heckman^{rank}$ treats selection as a binary variable, we propose two bias-correcting ensembles that account for the nuanced probability of being selected due to position bias and combine $Heckman^{rank}$ with existing position-bias correction methods. 

Our experimental evaluations demonstrate the utility of our proposed method when compared to state-of-the-art algorithms for unbiased learning-to-rank. Our ensemble methods have better accuracy compared to existing unbiased LTR algorithms under realistic selection bias assumptions, especially %
when the position bias is not severe. Moreover, $Heckman^{rank}$ is more robust to noise than both ensemble methods and position-bias correcting methods across difference position bias assumptions. The experiments also show that selection bias affects the performance of LTR systems even in the absence of position bias, and $Heckman^{rank}$ is able to correct for it.  %

\section{Related work}

Here, we provide the context for our work and present the three areas that best encompass our problem: bias in recommender systems, selection bias correction, and unbiased learning-to-rank.

\textbf{Bias in recommender system}.
Many technological platforms, such as recommendation systems, tailor items to users by filtering and ranking information according to user history. This process influences the way users interact with the system and how the data collected from users is fed back to the system and can lead to several types of biases. \citet{chaney-recsys18} explore a closely related problem called algorithmic confounding bias, where live systems are retrained to incorporate data that was influenced by the recommendation algorithm itself. Their study highlights the fact that training recommendation platforms with naive data that are not debiased can cause a severe decrease in the utility of such systems. For example, “echo chambers” are consequence of this problem \cite{fleder2007recommender, dan2013long}, where users are limited to an increasingly narrower choice set over time which can lead to a phenomenon called polarization \cite{dandekar2013biased}. Popularity bias, is another bias affecting recommender system that is studied by \citet{celma2008hits}. Popularity bias refers to the idea that a recommender system will display the most popular items to a user, even if they are not the most relevant to a user's query.
Recommender systems can also affect users decision making process, known as decision bias, and ~\citet{chen2013human} show how understanding this bias can improve recommender systems.
Position bias is yet another type of bias that is studied in the context of learning-to-rank systems and refers to documents that higher ranked will be more likely to be selected regardless of the document's relevancy.
\citet{joachims-wsdm17} focus on this bias and we compare our results to theirs throughout.

\textbf{Selection bias correction}. Selection bias occurs when a data sample is not representative of the underlying data distribution. Selection bias can have various underlying causes, such as participants self-selecting into a study based on certain criteria, or subjects choosing over a choice set that is restricted in a non-random way. Selection bias could also encompass the biases listed above.
Various studies attempt to correct for selection bias in different contexts.

Heckman correction, and more generally, bivariate selection models, control for the probability of being selected into the sample when predicting outcomes~\cite{heckman-econometrica79}.  
\citet{smith-kdd04} study Heckman correction for different types of selection bias through Bayesian networks, but not in the context in learning-to-rank systems. \citet{zadrozny2004learning} study selection bias in the context of well-known classifiers, where the outcome is binary rather than continuous as with ranking algorithms. 
Selection bias has also been studied in the context of causal graphs~\cite{bareinboim-aaai14,bareinboim-pnas16,correa-aaai18, bareinboim2015recovering, correa2017causal}. For example, if an underlying data generation model is assumed, ~\citet{bareinboim-aistats12} show that selection bias can be removed even in the presence of confounding bias, i.e., when a variable can affect both treatment and control. We leverage this work in our discussion of identifiability under selection bias. %

The most related work to our context are studies by
\citet{schnabel2016recommendations, wang2018deconfounded, hernandez2014probabilistic}.  
Both \citet{schnabel2016recommendations} and \citet{hernandez2014probabilistic} use a matrix factorization model to represent data (ratings by users) that are missing not-at-random, where \citet{schnabel2016recommendations} outperform \citet{hernandez2014probabilistic}. More recently, \citet{joachims-wsdm17} propose a position debiasing approach in the context of learning-to-rank systems as a more general approach compared to \citet{schnabel2016recommendations}. Throughout, we compare our results to \citet{joachims-wsdm17}, although, it should be noted that the latter deals with a more specific bias - position bias - than what we address here.
Finally, \citet{wang2018deconfounded} address selection bias due to confounding, whereas we address selection bias that is treatment-dependent only. %

\textbf{Unbiased learning-to-rank}.
The problem we study here investigates debiasing data in learning-to-rank systems. %
There are two approaches to LTR systems, offline and online, and the work we propose here falls in the category of offline LTR systems. 

Offline LTR systems learn a ranking model from historical click data and interpret clicks as absolute relevance indicators~\cite{joachims2005accurately, joachims-wsdm17, ai-sigir18, borisov2016neural, chapelle2009dynamic, craswell-wsdm08, joachims2002optimizing, richardson2007predicting, wang-wsdm18, wang-sigir16, hu2019unbiased}. Offline approaches must contend with the many biases that found data are subject to, including position and selection bias, among others. 
For example, \citet{wang-sigir16} use a propensity weighting approach to overcome position bias.
Similarly,~\citet{joachims-wsdm17} propose a method to correct for position bias, by augmenting $SVM^{rank}$ learning with an Inverse Propensity Score defined for clicks rather than queries. They demonstrate that Propensity-Weighted $SVM^{rank}$ outperforms a standard Ranking $SVM^{rank}$ by accounting for position bias. More recently \citet{agarwal2019general} proposed nDCG $SVM^{rank}$ that outperforms Propensity-Weighted $SVM^{rank}$~\cite{joachims-wsdm17}, but only when position bias \textit{is} severe. %
We show that our proposed algorithm outperforms ~\cite{joachims-wsdm17} when position bias \textit{is not} severe. Thus, we do not compare our results to \cite{agarwal2019general}.

Other studies aim to improve on~\citet{joachims-wsdm17}, such as~\citet{wang-wsdm18} and~\citet{ai-sigir18}, but only in the ease of their methodology. ~\citet{wang-wsdm18} propose a regression-based Expectation Maximization method for estimating the click position bias, and its main advantage over ~\citet{joachims-wsdm17} is that it does not require randomized tests to estimate the propensity model. Similarly, the Dual Learning Algorithm (DLA) proposed by~\citet{ai-sigir18} jointly learns the propensity model and ranking model without randomization tests.
~\citet{hu2019unbiased} introduce a method that jointly estimates position bias and trains a ranker using a pairwise loss function. The focus of these latter studies is position bias and not selection bias, namely the fact that some relevant documents may not be exposed to users at all, which is what we study here.

In contrast to offline LTR systems, online LTR algorithms intervene during click collection by interactively updating a ranking model after each user interaction~\cite{hofmann2013reusing, oosterhuis2018differentiable, schuth2014multileaved, yue2009interactively, chapelle2012large, raman2013learning, schuth2016multileave, jagerman2019model}. This can be costly, as it requires intervening with users' experience of the system. 
The main study in this context is \citet{jagerman2019model} who compare the online learning approach by \citet{oosterhuis2018differentiable} with the offline LTR approach proposed by ~\citet{joachims-wsdm17} under selection bias. The study shows that the method by ~\citet{oosterhuis2018differentiable} outperforms ~\cite{joachims-wsdm17} when selection bias and moderate position bias exist, and when no selection bias and severe position bias exist. One advantage of our offline algorithms over online LTR ones is that they do not have a negative impact on user experience while learning.

\section{Problem description}
\label{sec:problem}

In this section, we review the definition of learning-to-rank systems, position and selection bias in recommender systems, as well as our framing of bias-corrected ranking with counterfactuals. 

\subsection{Learning-to-Rank Systems}
We first describe learning-to-rank systems assuming knowledge of true relevances (full information setting) following \cite{joachims-wsdm17}. Given a sample $\mathbf{x}$ of i.i.d. queries ($\mathbf{x}_i \sim P(\mathbf{x})$) and relevancy score rel($\mathbf{x},y$) for all documents $y$, we denote $\Delta \ (\mathbf{y} | \mathbf{x}_i)$ to be the loss of any ranking $\mathbf{y}$ for query $\mathbf{x}_i$.
The risk of ranking system $S$ that returns ranking $S(\mathbf{x})$ for queries $\mathbf{x}$ is given by:
 \begin{align}
     R(S)= \int \Delta \ (S(\mathbf{x}) | \mathbf{x}) \ d \ P(\mathbf{x}).
     \label{103}
 \end{align}

Since the distribution of queries is not known in practice, $R(S)$ cannot be computed directly, it is often estimated empirically as follows:
 \begin{align}
     \hat{R}(S)= \frac{1}{\abs{\mathbf{X}}} \sum_{\mathbf{x}_i \in \mathbf{X}}  \Delta \ (S(\mathbf{x}_i) | \mathbf{x}_i) .
     \label{102}
 \end{align}
The goal of learning-to-rank systems is to find a ranking function $S \subset \mathcal{S}$ that minimizes the risk $\hat{R}(S)$. Learning-to-rank systems are a special case of a recommender system where, appropriate ranking is learned.

The relevancy score rel($\mathbf{x}_i,y$) denotes the true relevancy of document $y$ for a specific query $\mathbf{x}_i$. It is typically obtained via human annotation, and is necessary for the full information setting. Despite being reliable, true relevance assignments are frequently impossible or expensive to obtain because they require a manual evaluation of every possible document given a query.  %

Due to the cost of annotation, recommender system training often relies on implicit feedback from users in what is known as \textit{partial information setting}. Click logs collected from users are easily observable in any recommender system, and can serve as a proxy to the relevancy of a document. For this reason clicks are frequently used to train new recommendation algorithms. %
Unfortunately, there is a cost for using click log data because of noise (e.g., people can click on items that are not relevant) and various biases that the data are subject to, including position bias and selection bias which we discuss next.

\subsection{Position bias}
\label{subsec:position}

Implicit feedback (clicks) in LTR systems is inherently biased. %
Position bias refers to the notion that higher ranked results are more likely to be clicked by a user even if they are not the most relevant results given a user’s query.

Previous work \cite{joachims-wsdm17,wang-sigir16} has focused on tempering the effects of position bias via inverse propensity weighting (IPW). IPW re-weights the relevance of documents using a factor inversely related to the documents' position on a page.
For a given query instance $\mathbf{x}_i$, the relevance of document $y$ to query $\mathbf{x}_i$ is $r_i(y) \in \{0,1\}$, and $\mathbf{o}_i \in \{0,1\}$ is a set of vectors indicating whether a document $y$ is observed. Suppose the performance metric of interest is the sum of the rank of relevant documents:
 \begin{align}
     \Delta \ (\mathbf{y} | \mathbf{x}_i, r_i) = \sum_{y \in \mathbf{y}} rank (y | \mathbf{y} ) \ r_i(y).
     \label{100}
 \end{align}
Due to position bias, given a presented ranking $\mathbf{\Bar{y}}_i$, clicks are more likely to occur for top-ranked documents. Therefore, the goal is to obtain an unbiased estimate of $\Delta \ (\mathbf{y} | \mathbf{x}_i, r_i)$ for a new ranking $\mathbf{y}$.

There are existing approaches that address position bias in LTR systems. For example, \textit{Propensity $SVM^{rank}$}, proposed by \citet{joachims-wsdm17}, is one such algorithm. It uses inverse propensity weights (IPW) to counteract the effects of position bias:

 \begin{align}
     \hat{\Delta}_{IPW} \ (\mathbf{y} | \mathbf{x}_i, \mathbf{\Bar{y}}_i, o_i) = \sum_{y: o_i=1 \wedge r_i=1} \frac{rank (y | \mathbf{y})}{Q(o_i=1 | \mathbf{x}_i, \mathbf{\Bar{y}}_i, r_i)}
     \label{101}
 \end{align}
where the propensity weight $Q(o_i=1 | \mathbf{x}_i, \mathbf{\Bar{y}}_i, r_i)$ denotes the marginal probability of observing the relevance $r_i(y$) of result $y$ for query $\mathbf{x}_i$, when the user is presented with ranking $\mathbf{\Bar{y}_i}$. \citet{joachims-wsdm17} estimated the IPW to be:

\begin{align}
     Q(o_i=1 | \mathbf{x}_i, \mathbf{\Bar{y}}_i, r_i) = {\left(\frac{1}{rank(y | \mathbf{\Bar{y}}_i)}\right)}^{\eta}
     \label{101}
 \end{align}
where $\eta$ is severity of position bias. The IPW has two main properties. First, it is computed only for documents that are observed and clicked. Therefore, documents that are never clicked do not contribute to the IPW calculation. Second, as shown by Joachims et al. \cite{joachims-wsdm17}, a ranking model trained with clicks and the IPW method will converge to a model trained with true relevance labels, rendering a LTR framework robust to position bias.

\subsection{Selection bias}
LTR systems rely on implicit feedback (clicks) to improve their performance. However, a sample of relevant documents from click data does not reflect the true distribution of all possible relevant documents because a user observes a limited choice of documents. This can occur because i) a recommender system ranks relevant documents too low for a user to feasibly see, or ii) because a user can examine only a truncated list of top $k$ recommended items. As a result, clicked documents are not randomly selected for LTR systems to be trained on, and therefore cannot reveal the relevancy of documents that were excluded from the ranking 
$\mathbf{\Bar{y}}$. This leads to selection bias. %

Selection bias and position bias are closely related. Besides selection bias due to unobserved relevant documents, selection bias can also arise due to position bias: lower ranked results are less likely to be observed, and thus selected more frequently than higher-ranked ones. %
Previous work on LTR algorithms that corrects for position bias assigns a non-zero observation probability to all documents, and proofs of debiasing are based on this assumption~\cite{joachims-wsdm17}. However, in practice it is rarely realistic to assume that all documents can be observed by a user. When there is a large list of potentially relevant documents, the system may choose to show only the top $k$ results and a user can only act on these results. Therefore, lower-ranked results are never observed, which leads to selection bias. Here, we consider the selection bias that arises when some documents have a zero probability of being observed if they are ranked below a certain cutoff $k$. %
The objective of this paper is to propose a ranking algorithm that corrects for 
both selection and position bias, and therefore is a better tool for training future LTR systems (see Section \ref{sec:heckman-solution}).

\subsection{Ranking with counterfactuals}
We define the problem of ranking documents as a counterfactual problem~\cite{pearl2016causal}.
Let 
$O(\mathbf{x},y) \in \{0,1\}$ denote a treatment variable indicating whether a user observed document $y$ given query $\mathbf{x}$. Let $C_{O=1}(\mathbf{x},y) \in \{0,1\}$ %
represent the \textit{click counterfactual} indicating whether a document $y$ would have been clicked had $y$ been observed under query $\mathbf{x}$. %
The goal of \textit{ranking with counterfactuals} is to reliably estimate the probability of click counterfactuals for %
all documents: %
 \begin{align}
       P(C_{O=1}=1|\mathbf{X=x},Y=y)
 \end{align}
and then rank the documents according to this probability. Solving the ranking with counterfactuals problem would allow us to find a ranking system $S$ that returns ranking $S(\mathbf{x})$ for query $\mathbf{x}$ that is robust to selection bias.

Current techniques that correct for position bias aim to provide reliable estimates of this probability by taking into consideration the rank-dependent probability of being observed. However, this approach is only effective for documents that have a non-zero probability of being observed:
 \begin{align}
 P(C_{O=1}=1|O=1, rank=i,\mathbf{X=x},Y=y).
     \label{eq:biased}   
 \end{align}
The challenge with selection bias is to estimate this probability for documents that have neither been observed nor clicked in the first place:

 \begin{align}
       P(C_{O=1}=1|O=0,C=0,\mathbf{X=x},Y=y)\\=P(C_{O=1}=1|O=0,\mathbf{X=x},Y=y).
     \label{eq:counterfactual}   
 \end{align}
To address this challenge, in the following Section~\ref{sec:solution} we turn to econometric methods, which have a long history of addressing selection bias.

Note that in order to recover from selection bias we must address the concept of identifiability and whether causal estimates can even be obtained in the context of our setup. A bias is identified to be recoverable if the treatment is known~\cite{bareinboim-aaai14}. In our context the treatment is whether a document enters into the data training pool (clicked). While it is difficult to guarantee that a user observed a document that was shown to them (i.e. we cannot know whether an absence of a click is due non-observance or to non-relevance), it is easier to guarantee that a document was not observed by a user if it was not shown to them in the first place (e.g., it is below a cutoff for top  %
$k$ results or the user never scrolled down to that document in a ranked list). Our proposed solution, therefore, identifies the treatment first as a binary variable (whether the document is shown versus not shown) and then as a continuous variable that takes position bias into account.

\section{Bias-corrected ranking with counterfactuals} 
\label{sec:solution}

In this section we adapt a well-known sample selection correction method, known as Heckman's two-stage correction, to the context of LTR systems. Integrating the latter framework requires a detailed understanding of how LTR systems generally process and cut interaction data to train new recommendation algorithms, and at what stages in that process selection biases are introduced. Thus, while the methodology we introduce is a well established tool in the causal inference literature, integrating it within the multiple stages of training a machine learning algorithm is a complex translational problem.  
We then introduce two aggregation methods to combine our proposed $Heckman^{rank}$, correcting for selection bias, with existing methods for position bias to further improve the accuracy in ranking prediction.  

\subsection{Selection bias correction with $Heckman^{rank}$} \label{sec:heckman-solution}
Econometrics, or the application of a statistical methods to economic problems, has long been concerned with confounded or held-out data in the context of consumer choices. Economists are interested in estimating models of consumer choice to both learn consumers' preferences and to predict their outcomes. Frequently, the data used to estimate these models are observational, not experimental. As such, the outcomes observed in the data are based on a limited and self-selected sample. A quintessential example of this problem is estimating the impact of education on worker's wages based on only those workers who are employed \cite{heckman-econometrica79}. However, those who are employed are a self-selected sample, and estimates of education's effect on wages will be biased.

A broad class of models in econometrics that deal with such selection biases are known as bivariate sample selection models. A well-known method for correcting these biases in economics is known as Heckman correction or two-step Heckman. In the first stage the probability of self selection is estimated, and in the second stage the latter probability is accounted for. As \citet{heckman-econometrica79} pointed out self selection bias can occur for two reasons. ``First, there may be self selection by the individuals or data units being investigated (as in the labor example). Second, sample selection decisions by analysts or data processors operate in much the same fashion as self selection (by individuals).''

Adapting a sample selection model, such as Heckman's, to LTR systems requires an understanding of when and how data are progressively truncated when training a recommender algorithm. We introduce notation and a framework to outline this problem here. 

Let  $c_{x,y}$ denote whether a document $y$ is selected (e.g., clicked) under query $x$ for each $<query, document>$ pair; $\bm{F}_{x,y}$ represents the features of the $<query, document>$, and $\epsilon_{x,y}$ is a normally distributed error term. The same query can produce multiple $<query, document>$ pairs, where the documents are then ordered by a LTR algorithm. However, it is important to note that a LTR algorithm will not rank every single document in the data given a $query$. Unranked documents are typically discarded when training future algorithms. Herein lies the selection bias. Documents that are not shown to the user can then never be predicted as a potential choice. Moreover, documents far down in the rankings may still be kept in future training data, but will appear infrequently. Both these points will contribute to generating increasingly restrictive data that new algorithms are trained on.

If we fail to account for the repercussions of these selection biases, then modeling whether a document is selected will be based only upon the features of documents that were ranked and shown to the user, which can be written as: 
\begin{align}
c_{x,y} = \bm{\alpha}^{biased} \bm{F}_{x,y} + \epsilon_{x,y}.
\label{1}
\end{align}

In this setup we only consider a simple linear model; however, future research will incorporate nonlinear models. In estimating (\ref{1}), we refer to the feature weights estimator, $\bm{\alpha}^{biased}$, as being biased, because the feature design matrix will only reflect documents that were shown to the user. But documents that were not shown to the user could also have been selected. Thus, (\ref{1}) reflects the limitation outlined in (\ref{eq:biased}). When we discard unseen documents then we can only predict clicks for documents that were shown, while our objective is to predict the unconditional probability that a document is clicked regardless of whether it was shown.

To address this point, we will first explicitly model an algorithm's document selection process. Let $o_{x,y}$ denote a binary variable that indicates whether a document $y$ is shown and observed ($o_{x,y} = 1$) or not shown and not observed ($o_{x,y} = 0$). For now, we assume that if a document is shown to the user that user also sees the document. We relax this assumption in Section ~\ref{sec:combo-solution}.  $\bm{Z}_{x,y}$ is a set of explanatory variables that determine whether a document is shown, which includes the features in $\bm{F}_{x,y}$, but can also include external features, including characteristics of the algorithm that first generated the data: %
\begin{align}
o_{x,y} = \mathds{1}_{\{\bm{\theta}\bm{Z}_{x,y} + \epsilon^{(1)}_{x,y}\geq 0\}}.
\label{2}
\end{align}

In the first stage of $Heckman^{rank}$, we estimate the probability of a document being observed using a Probit model:
\begin{align}
P(o_{x,y} = 1|\bm{Z}_{x,y}) = P( \bm{\theta}\bm{Z}_{x,y} + \epsilon_{x,y} > 0| \bm{Z}_{x,y})=\Phi( \bm{\theta} \bm{Z}_{x,y})
\label{3}
\end{align}
where $\Phi()$ denotes the standard normal CDF. Note that a crucial assumption here is that we will use both seen and unseen documents for a given $query$ in estimating (\ref{3}). Therefore, the dimensions of our data will be far larger than if we had discarded unseen documents, as most LTR systems typically do. After estimating (\ref{3}) we can compute what is known as an Inverse Mills ratio for every $<query, document>$ pair: 
\begin{align}
\lambda_{x,y} = \frac{\phi(\bm{\theta} \bm{Z}_{x,y})}{\Phi( \bm{\theta} \bm{Z}_{x,y})}
\label{eq:mills}
\end{align}
where $\phi()$ is the standard normal distribution. $\lambda_{x,y}$ reflects the 
severity of selection bias and corresponds to our desire to condition on $O = 0$ versus $O = 1$, as described in Equations \ref{eq:biased} and \ref{eq:counterfactual}, but using a continuous variable reflecting the probability of selection. 

In the second stage of $Heckman^{rank}$, we estimate the probability of whether a user will click on a document. Heckman's seminal work showed that if we condition our estimates on the $\lambda_{x,y}$ our estimated feature weights will be statistically unbiased in expectation. This can improve our predictions if we believe that including $\lambda_{x,y}$ is relevant in predicting clicks. 
We assume joint normality of the errors, and our setup naturally implies that the error terms $\epsilon_{x,y}$ and $\epsilon^{(1)}_{x,y}$ are correlated, namely that clicking on a document depends upon whether a document is observed by users and, therefore, has the potential for being selected.

The conditional expectation of clicking on a document conditional on the document being shown is given by:
\begin{align}
\mathbb{E}[c_{x,y} | F, o_{x,y} = 1] = \bm{\alpha}\bm{F}_{x,y} + \mathbb{E}(\epsilon_{x,y}| F, o_{x,y}=1) 
 = \bm{\alpha}\bm{F}_{x,y} + \sigma \bm{\lambda}_{x,y} 
\label{4}
\end{align}

We can see that if the error terms in (\ref{1}) and (\ref{2}) are correlated then $\mathbb{E}(\epsilon_{x,y}| F, o_{x,y}=1) > 0 )$, and estimating (\ref{4}) without accounting for this correlation will lead to biased estimates of $\bm{\alpha}$. 
Thus, in the second stage, we correct for selection bias to obtain an unbiased estimate of $\bm{\alpha}$ by controlling for $\hat{\bm{\lambda}}_{x,y}$:
\begin{align}
c_{x,y} =  \bm{\alpha}^{unbiased}\bm{F}_{x,y} + \sigma \hat{\bm{\lambda}}_{x,y}(\hat{\bm{\theta}}\bm{Z}_{x,y}) + \epsilon_{x,y} 
\label{5}
\end{align}

Estimation of (\ref{5}) allows us to predict click probabilities, $\hat{c}$, where $\hat{c} = \bm{\hat{\alpha}}^{unbiased}\bm{F}_{x,y} + \hat{\sigma} \hat{\bm{\lambda}}_{x,y}(\hat{\bm{\theta}}\bm{Z}_{x,y}) $. This click probability refers to our ability to estimate (\ref{eq:counterfactual}), the unconditional click probability, using $Heckman^{rank}$. We then compute document rankings for a given query by sorting documents according to their predicted click probabilities. %
Note that our main equation (\ref{5}) has a bivariate outcome. Thus, in this selection correction setup we are following a Heckprobit model, as opposed to the initial model that Heckman proposed in \citet{heckman-econometrica79} where the main outcome is a continuous variable.

Our setup helps account for the inherent selection bias that can occur in any LTR system, as all LTR systems must make a choice in what documents they show to a user. What is unique to our formulation of the problem is our use of a two stage estimation process to account for the two stage document selection process: namely, whether the document is shown, and whether the document is then selected. Accounting for the truncation of the data is critical for training a LTR system, and previously has not been considered. In order to improve a system's ranking accuracy it must be able to predict document selection for both unseen as well as seen documents. If not, the choice set of documents that are available to a user can only become progressively smaller. Our correction is a simple method to counteract such a trend in found data.

\subsection{Bias-correcting ensembles} \label{sec:combo-solution}

Biased data limits the ability to accurately train an LTR algorithm on click logs. In this section, we present methods for addressing two types of selection bias, one stemming from truncated recommendations and the other one from position bias. One of the deficiencies of using $Heckman^{rank}$ to deal with biased data is that it assumes that all documents that are shown to a user are also observed by the user. However, due to position bias that is not necessarily the case, and lower-ranked shown documents have lower probability of being observed. Therefore, it is natural to consider combining $Heckman^{rank}$, which focuses on recovering from selection bias due to unobserved documents, with a ranking algorithm that accounts for the nuanced observation probability of shown documents due to position bias.

Algorithms that rely on IPW~\cite{wang-sigir16, joachims-wsdm17, agarwal2019general} consider the propensity of observation for any document given a ranking for a certain query and it is exponentially dependent on the rank of the document in the given ranking. This is clearly different from our approach for recovering from selection bias where we model the observation probability to be either $0$ or $1$ depending on its position relative in the ranking.

\textbf{Ensemble ranking objective}. In order to harness the power of correcting for these biases in a collective manner, we propose to use ensembles that can combine the results produced by \textit{$Heckman^{rank}$} and any position bias correcting method. We refer to the algorithm correcting for selection bias as $A_s$ and for position bias as $A_p$ while $\mathbf{y}_{s}$ and $\mathbf{y}_{p}$ are the rankings generated for a certain query $\mathbf{x}$ by the algorithms respectively. Our goal is to produce an ensemble ranking $\mathbf{y}_{e}$ based on $\mathbf{y}_{s}$ and $\mathbf{y}_{p}$ for all queries that is more accurate than either ranking alone. 

There is a wide scope for designing an appropriate ensemble method to serve our objective. We propose two simple but powerful approaches, as our experimental evaluation shows. The two approaches differ in their fundamental intuition. The intuition behind the first approach is to model the value of individual ranking algorithms through a linear combination of the rankings they produce. We can learn the coefficients of that linear combination using linear models on the training data. We call this method \textit{Linear Combination}. The second approach is a standard approach for combining ranking algorithms using Borda counts ~\cite{dwork-www01}. It works as a post processing step after the candidate algorithms $A_s$ and $A_p$ produce their respective rankings ${\mathbf{y}}_{s}$ and ${\mathbf{y}}_{p}$. We apply a certain \textit{Rank Aggregation} algorithm over $\mathbf{y}_{s}$ and $\mathbf{y}_{p}$ to produce $\mathbf{y}_{e}$ for a given query for evaluation. Next, we discuss each of the approaches in the context of our problem.

\subsubsection{\textbf{Linear Combination}}
A simple aggregation method for combining $A_s$ and $A_p$ is to estimate the value of each algorithm in predicting a click. %
After training the algorithms $A_s$ and $A_p$, we use the same training data to learn the weights of a linear model that considers the rank of each document produced by $A_s$ and $A_p$. For any given query $\mathbf{x}$ the ranking of document $y$ produced by $A_s$ is given by $rank(y|\mathbf{y}_s, \mathbf{x})$. Similarly, $rank(y|\mathbf{y}_p, \mathbf{x})$ represents the ranking given by $A_p$. We also consider the relevance of document $y$, $rel(\mathbf{x}, y)$ which is either $0$ for not relevant or $1$ for relevant, modeled through clicks.

We train a binary classifier to predict relevance (click) of documents which incorporates the estimated value of individual algorithms. We select logistic regression to be the binary classifier in our implementation, but any other standard classification method should work as well. We model the relevance of a document $y$, given two rankings $\mathbf{y}_s$ and $\mathbf{y}_p$ as the following logistic function:

\[
    rel(\mathbf{x}, y) = \frac{1}{1 + e^{-(w_0 + w_1 * rank(y|\mathbf{y_s},\mathbf{x}) + w_2 * rank(y|\mathbf{y_p},\mathbf{x}))}}
\]

Upon training the logistic regression model we learn the parameters $w_0, w_1, w_2$ where $w_1$ and $w_2$ represent the estimated impact of $A_s$ and $A_p$ respectively. During evaluation we predict the click counterfactual probability for each $< query, document >$ pair using the trained classifier. Then we can sort the documents for each query according to these probability values to generate the final ensemble ranking $\mathbf{y}_e$.

\subsubsection{\textbf{Rank Aggregation}}
Rank aggregation aims to combine rankings generated by multiple ranking algorithms. In a typical rank aggregation problem, we are given a set of rankings $\mathbf{y}_1, \mathbf{y}_2, ... , \mathbf{y}_m$ of a set of objects $y_1, y_2, ... , y_n$ given a query $\mathbf{x}$. The objective is to find a single ranking $\mathbf{y}$ that corroborates with all other existing rankings. Many aggregation methods have been proposed \cite{lin-wir2010}. A commonly used approach is the Borda count, which scores documents based on their relative position in a ranking, and then totals all scores across all rankings for a given query \cite{dwork-www01}. 

In our scenario, we have two rankings $\mathbf{y}_{s}$ and $\mathbf{y}_{p}$. For a given query, there are $n$ documents to rank. Consider $B(\mathbf{y}_{s}, y_i)$ as the score for document $y_i$ ($i \in \{1, 2, ... , n\}$) given by $\mathbf{y}_{s}$. Similarly, $B(\mathbf{y}_{p}, y_i)$ refers to the score for document $y_i$ given by $\mathbf{y}_{p}$. The total score for document $y_i$ would be $B(\mathbf{y}_{s}, y_i) + B(\mathbf{y}_{p}, y_i)$. Based on these total scores we sort the documents in non-ascending order of their scores to produce the ensemble ranking $\mathbf{y}_e$. The score given to a document $y_i$ in a specific ranking $\mathbf{y}_s$ (or $\mathbf{y}_p$) is simply the number of documents it beats in the respective ranking. For example, given a certain query, if a document is ranked $1^{st}$ in $\mathbf{y}_s$ and $3^{rd}$ in $\mathbf{y}_p$ then the total score for this document would be $(n-1) + (n-3)$. This very simple scheme reflects the power of the combined method to recover from different biases in LTR systems.

\section{Experiments}
\label{sec:experiments}

In this section, we evaluate our proposed approach for addressing selection bias %
under several conditions:
\begin{itemize}
\item Varying the number of observed documents given a fixed position bias (Section~\ref{sec:Experimental results})
\item Varying position bias with no noise (Section~\ref{subsec.1})
\item Varying position bias with noisy clicks (Section~\ref{subs:noise})
\item Varying noise level in click sampling (Section~\ref{subs:vary.noise})
\end{itemize}
The parameter values are summarized in Table \ref{table1}.

\subsection{Experimental setup}
\label{subsec:Experimental setup}

Next, we describe the dataset we use, the process for click data generation, and the evaluation framework.

\subsubsection{\textbf{Base dataset}}
\label{subsec: Dataset}
In order to explore selection bias in LTR systems, we conduct several experiments using semi-synthetic datasets based on set 1 and set 2 from the
Yahoo! Learning to Rank Challenge (C14B)~\footnote{\url{https://webscope.sandbox.yahoo.com/catalog.php?datatype=c}}, denoted as $D_{YLTR}$.
Set 1 contains %
$19,944$ train and $6,983$ test queries including $473,134$ train and $165,660$ test documents. Set 2 contains %
$1,266$ train queries and $34,815$ train documents, with $20$ documents per query on average~\cite{chapelle-11yahoo}.
Each query is represented by an $id$ and each <query, document> pair is represented by a $700$-dimensional feature vector with normalized feature values $\in [0,1]$. The dataset contains true relevance of rankings based on expert annotated relevance score $\in [0,4]$ associated with each $<query, document>$ pair, with $0$ meaning least relevant and $4$ most relevant. We binarized the relevance score following \citet{joachims-wsdm17}, such that $0$ denotes irrelevant (a relevance score of $0$, $1$ or $2$), and $1$ relevant (a score of $3$ and $4$). 

We first conduct extensive experiments on the train portion of the smaller set 2, where we randomly sample $70\%$ of the queries as training data and $30\%$ as test data, with which LTR algorithms can be trained and evaluated respectively (Section ~\ref{sec:Experimental results}).
To confirm the performance of our proposed method with out-of-sample test data, we conduct experiments on the larger set 1, where we train LTR algorithms on set 1 train data and evaluate them on set 1 test data (Section ~\ref{sec:Experimental results large}).

\subsubsection{\textbf{Semi-synthetic data generation}}
\label{subsec:Data generation algorithm}

We use the real-world base dataset, $D_{YLTR}$, to generate semi-synthetic datasets that contain document clicks for $<query, document>$ rankings. The main motivation behind using the Yahoo! Learning To Rank dataset is that it provides unbiased ground truth for relevant results, thus enabling unbiased evaluation of ranking algorithms. In real-world scenarios, unbiased ground truth is hard to come by and LTR algorithms are typically trained on biased, click data which does not allow for unbiased evaluation. To mimic real-world scenarios for LTR, the synthetic data generation creates such biased click data.

We follow the data-generation process of ~\citet{joachims-wsdm17}.
We train a base ranker, in our case $SVM^{rank}$, with $1\%$ of the training dataset that contains true relevances, and then use the trained model to generate rankings for the remaining $99\%$ of the queries in the training dataset. %
The second step of the data-generation process generates clicks on the ranked documents in the training dataset. The click probability of document $y$ for a given query $\mathbf{x}$ is calculated as
    $P(c_{x,y} = 1) = \frac{r_{i}(y)}{(rank(y|\mathbf{\Bar{y}))}^\eta}$
where $c_{x,y}(y)$ and $r_{i}(y)$ represent if a document $y$ is clicked and relevant respectively, $rank(y|\mathbf{\Bar{y}})$ denotes the ranking of document $y$ for query $\bm{x}$ if the user was presented the ranking $\mathbf{\Bar{y}}$, and $\eta$ indicates the severity of position bias.
Note that clicks are not generated for documents that are bellow a certain rank cutoff $k$ to incorporate the selection bias. 

In a single pass over the entire training data we generate clicks following the above click probability. We refer to this as one sampling pass. 
For the smaller set 2, we generate clicks over $15$ sampling passes, while for the larger set 1, we generate clicks over $5$ sampling passes.
This click-generation process reflects a common user behavior where some relevant documents do not receive any clicks, and other relevant documents receive multiple clicks. This process captures the generation of \textit{noiseless} clicks, where users only click on relevant documents. We also consider a click generation process with \textit{noisy} clicks in which a small percentage of clicks ($10-30\%$) occur on irrelevant documents.

\subsubsection{\textbf{Evaluation}}
\label{subsec:learning algorithm}

We explore the performance of LTR algorithms \textit{Naive $SVM^{rank}$},  \textit{Propensity  $SVM^{rank}$}, \textit{$Heckman^{rank}$} along with the two ensemble methods \textit{Linear Combination} (CombinedW) and \textit{Rank Aggregation} (RankAgg) with two different metrics: \textit{Average Rank of Relevant Results}
         $ARRR= \frac{\sum_{y: o_i=1 \wedge r_i=1} rank (y | \mathbf{\Bar{y}})}{\abs{\mathbf{X}}}$
and \textit{Normalized Discounted Cumulative Gain}
         ${nDCG}@p = {nDCG}_p) = \frac{{DCG}@p}{{IDCG}@p}$ 
where $p$ is the rank position up to which we are interested to evaluate, ${DCG}@p$ represents the discounted cumulative gain of the given ranking whereas ${IDCG}@p$ refers to the ideal discounted cumulative gain. We can compute ${DCG}@p$ using the following formula
         ${DCG}@p = \sum_{i=1}^{p} \frac{2^{rel(\mathbf{x}, y)} - 1}{log_2 \left(i + 1\right)}$
Similarly, %
         ${IDCG}@p = \sum_{i=1}^{|REL@p|} \frac{2^{rel(\mathbf{x}, y)} - 1}{log_2 \left(i + 1\right)}$
where $REL@p$ represents the list of relevant documents (ordered by their relevance) in the ranking up to position $p$ for a given query. In our evaluation we chose $p = 10$ for nDCG metric and we refer to it by nDCG@10. 

Each figure in the experiments depicts how the ARRR or nDCG@10 ($y$ axis) changes when the user only observes the first $k\in [1,30]$ documents ($x$ axis). Note that $k$ reflects the severity of selection bias as we model selection bias by assigning a zero observation probability to documents below cutoff $k$. In contrast, 
position bias is modeled by assigning a non-zero probability to every single document where $\eta$ represents the severity of the position bias. We vary severity of both selection bias and position bias with or without the existence of noise in click generation.
\scriptsize
\begin{table}
\begin{center}
 \begin{tabular}{|c|c|c|c|} 
 \hline
 parameter & value/category & description  & section\\ [0.5ex] 
 \hline
 $k$ & 1-30 & number of observed docs (selection bias) & \ref{sec:Experimental results}\\
 \hline
 $\eta$ & 0, 0.5, 1, 1.5, 2 & position bias severity & \ref{subsec.1}\\ 
 \hline
 noise & 0\%, 10\%, 20\%, 30\% & clicks on irrelevant docs & \ref{subs:noise}, \ref{subs:vary.noise} \\  
 \hline
\end{tabular}
\vspace{0.2cm}
\caption{Experimental Parameters}
\label{table1}
\vspace{-1.1cm}
\end{center}
\end{table}
\normalsize

\begin{figure*}
    \centering
    \subfloat[ARRR, $\eta = 0$]{\label{sfig:bias_a}\includegraphics[width=.25\textwidth, height=0.21\textwidth]{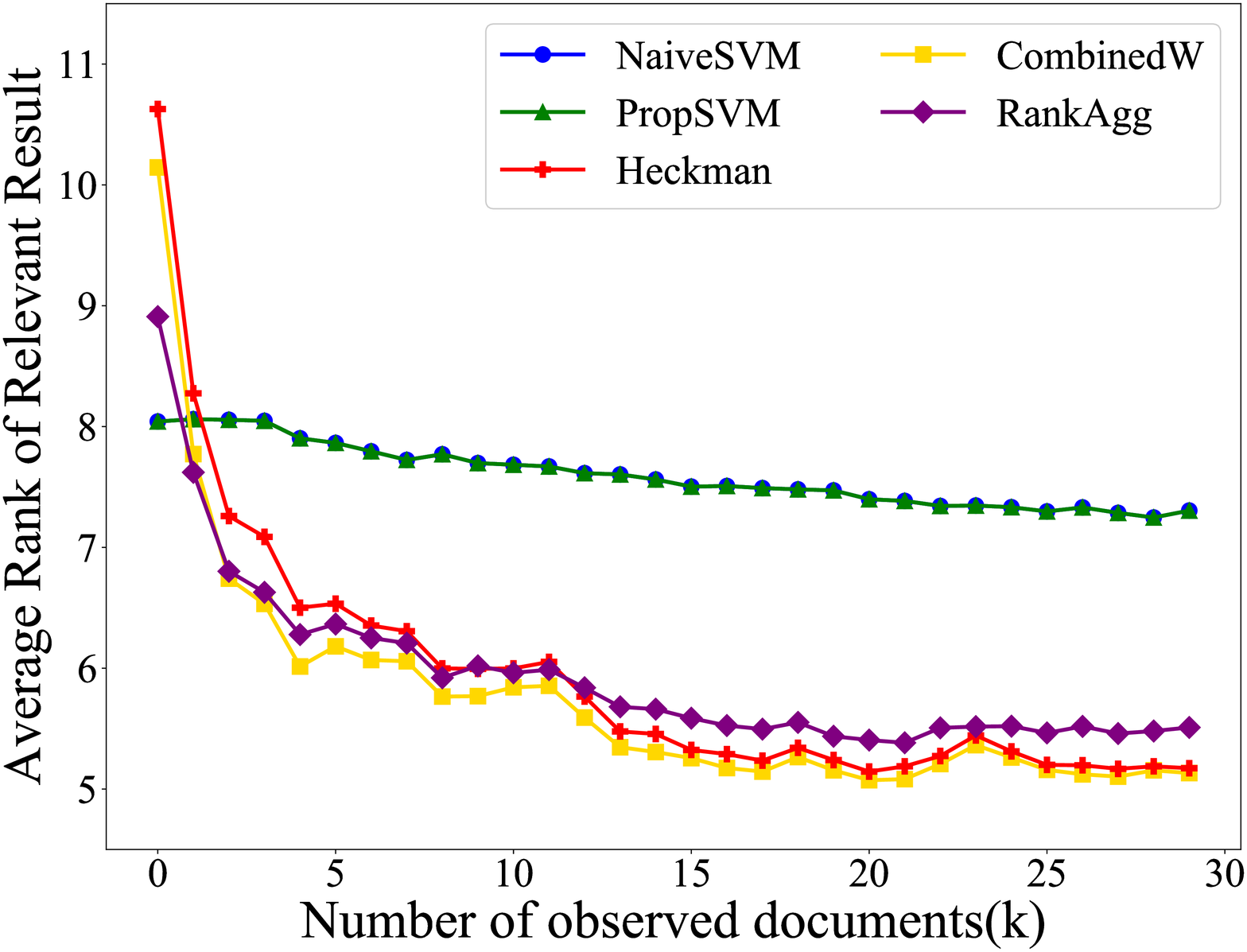}}\hfill
    \subfloat[ARRR, $\eta = 0.5$]{\label{sfig:bias_b}\includegraphics[width=.25\textwidth, height=0.21\textwidth]{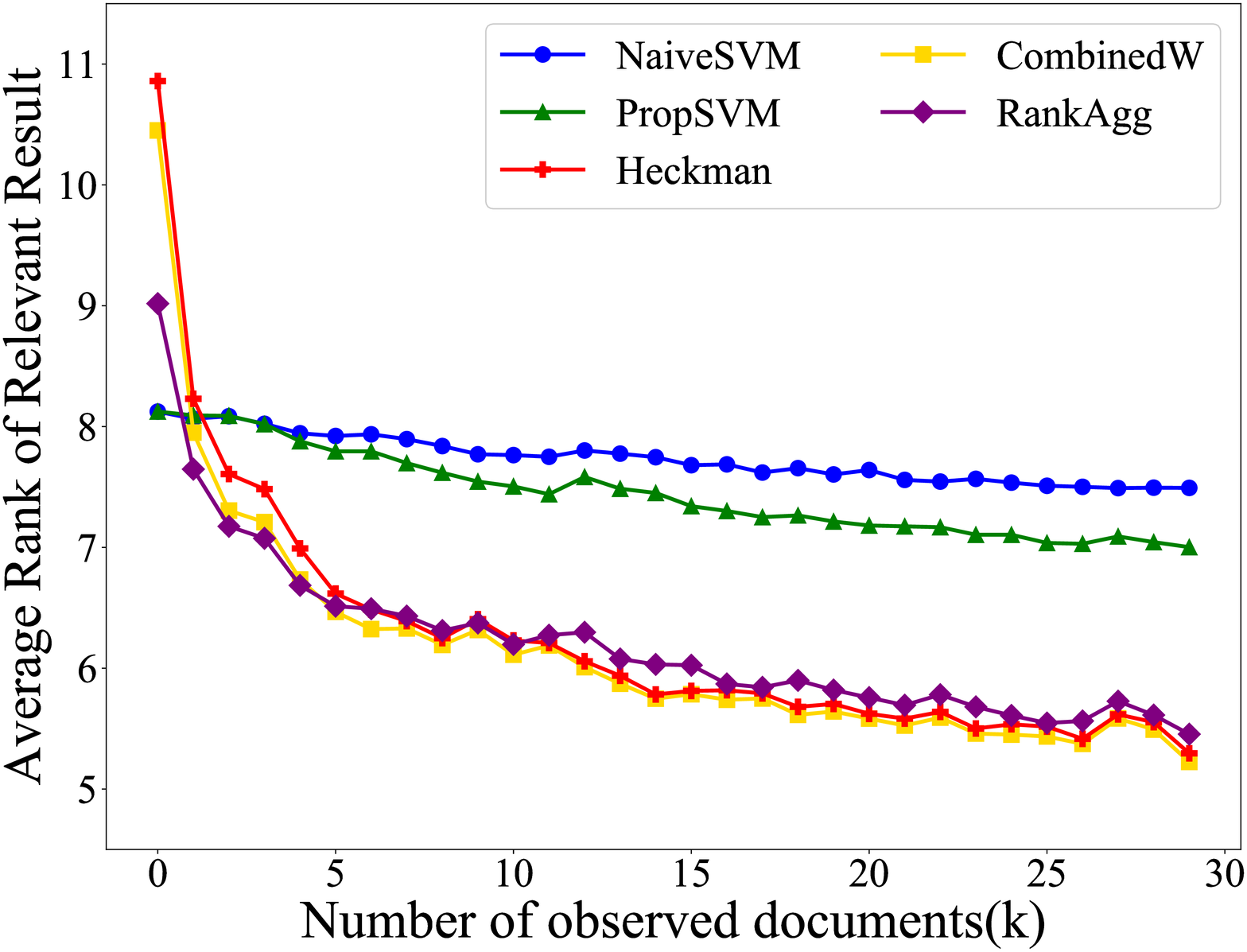}}\hfill
    \subfloat[ARRR, $\eta = 1$]{\label{sfig:bias_c}\includegraphics[width=.25\textwidth, height=0.21\textwidth]{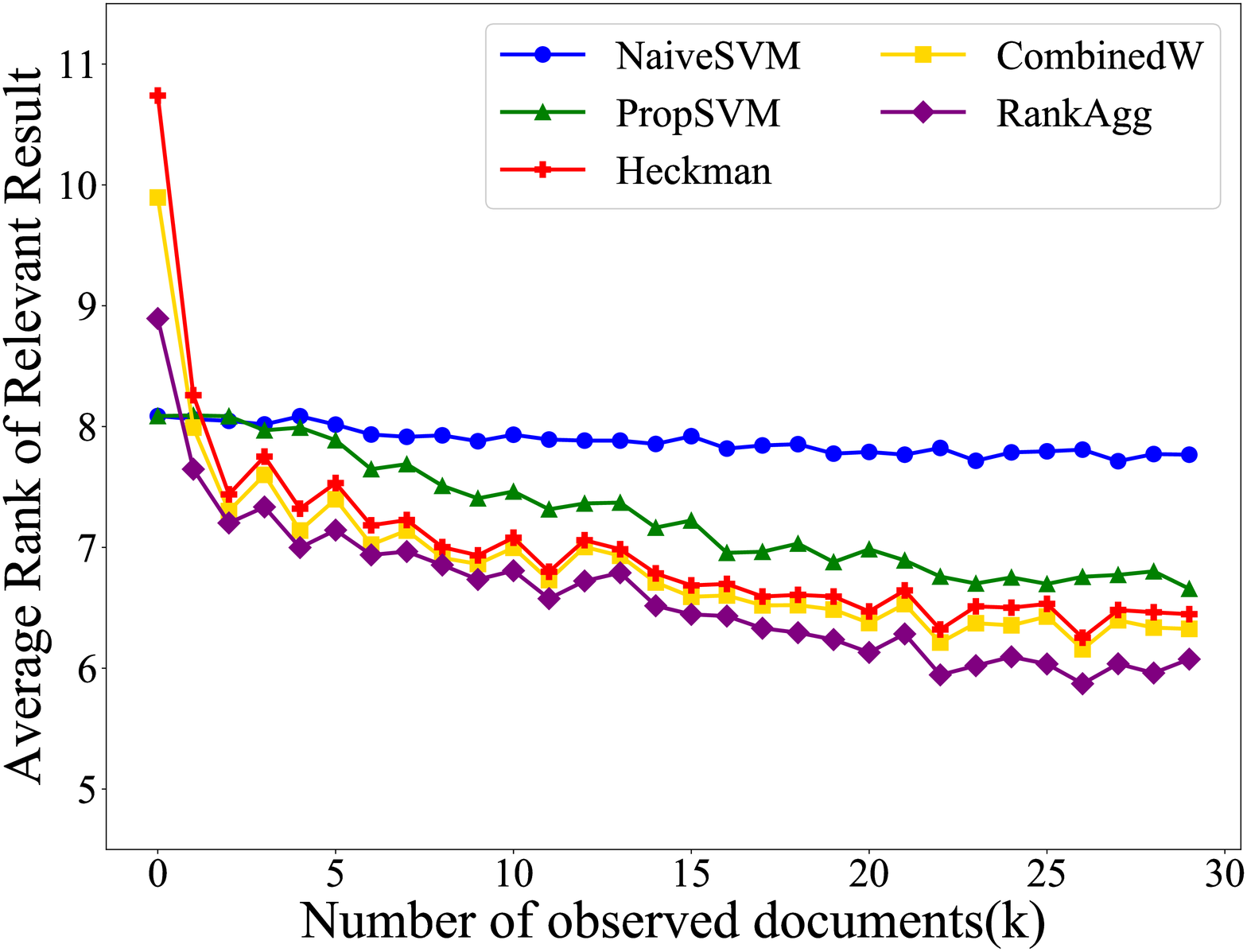}}\hfill
    \subfloat[ARRR, $\eta = 1.5$]{\label{sfig:bias_d}\includegraphics[width=.25\textwidth, height=0.21\textwidth]{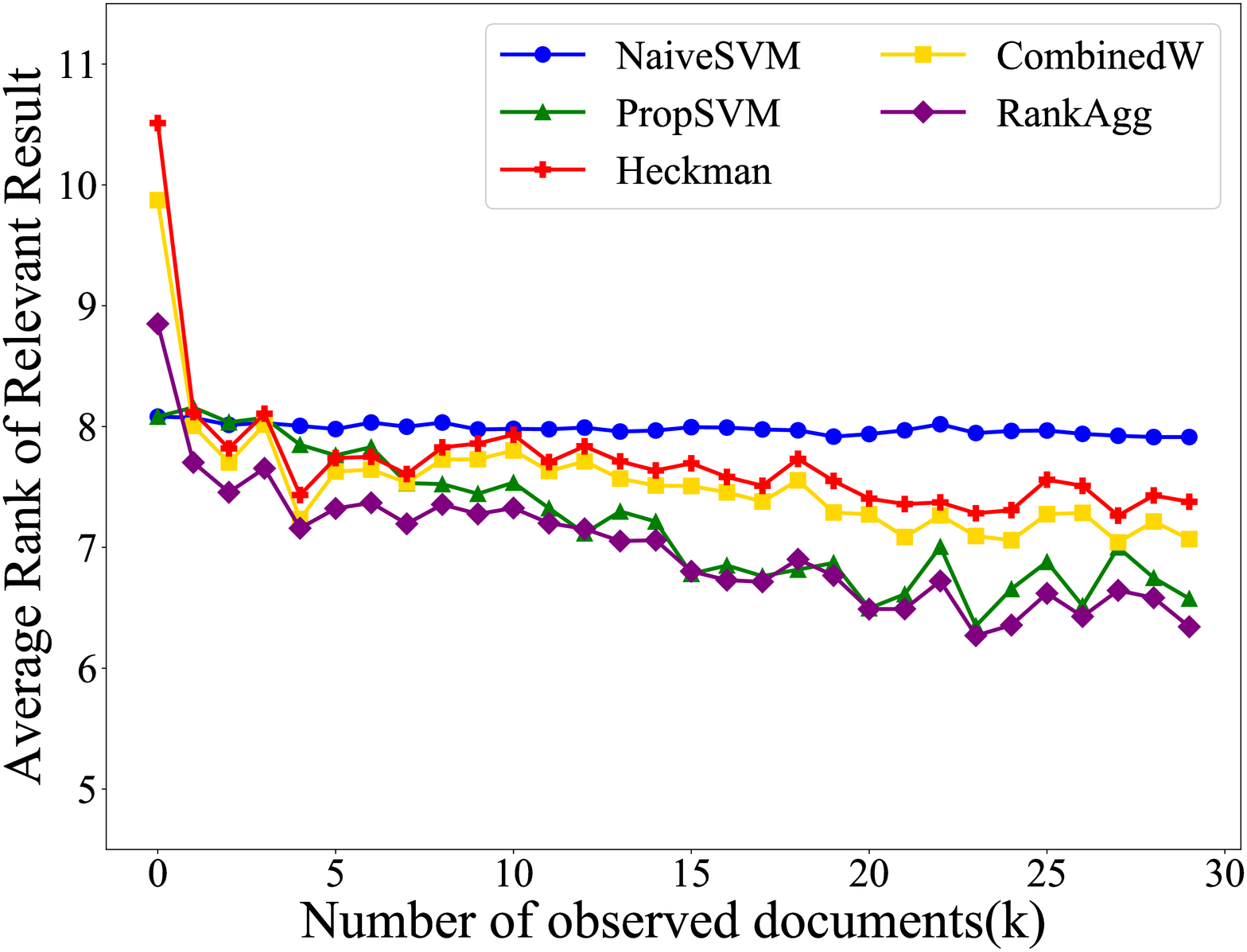}}\\
    
    \subfloat[nDCG@10, $\eta = 0$]{\label{sfig:bias_e}\includegraphics[width=.25\textwidth, height=0.21\textwidth]{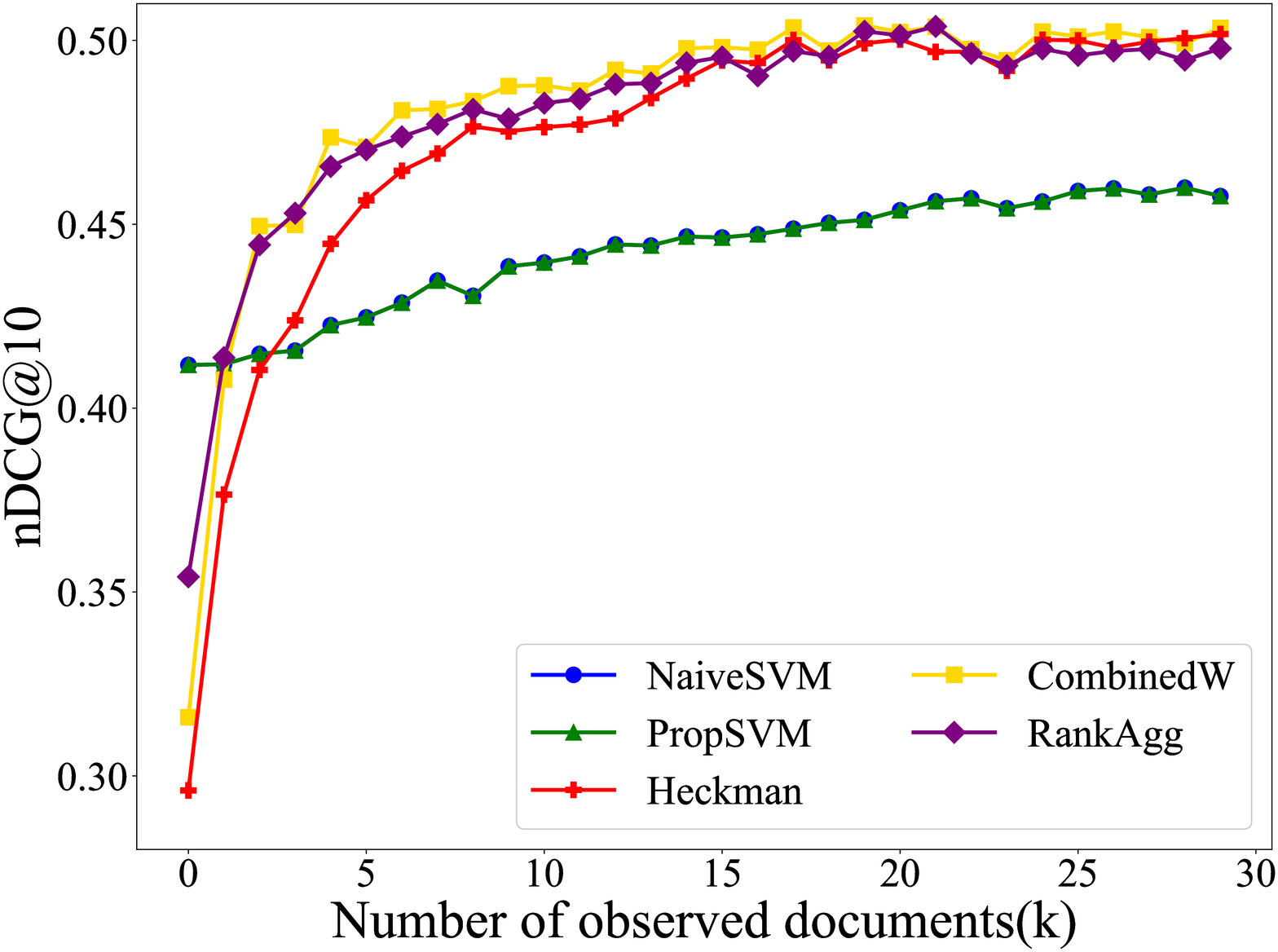}}\hfill
    \subfloat[nDCG@10, $\eta = 0.5$]{\label{sfig:bias_f}\includegraphics[width=.25\textwidth, height=0.21\textwidth]{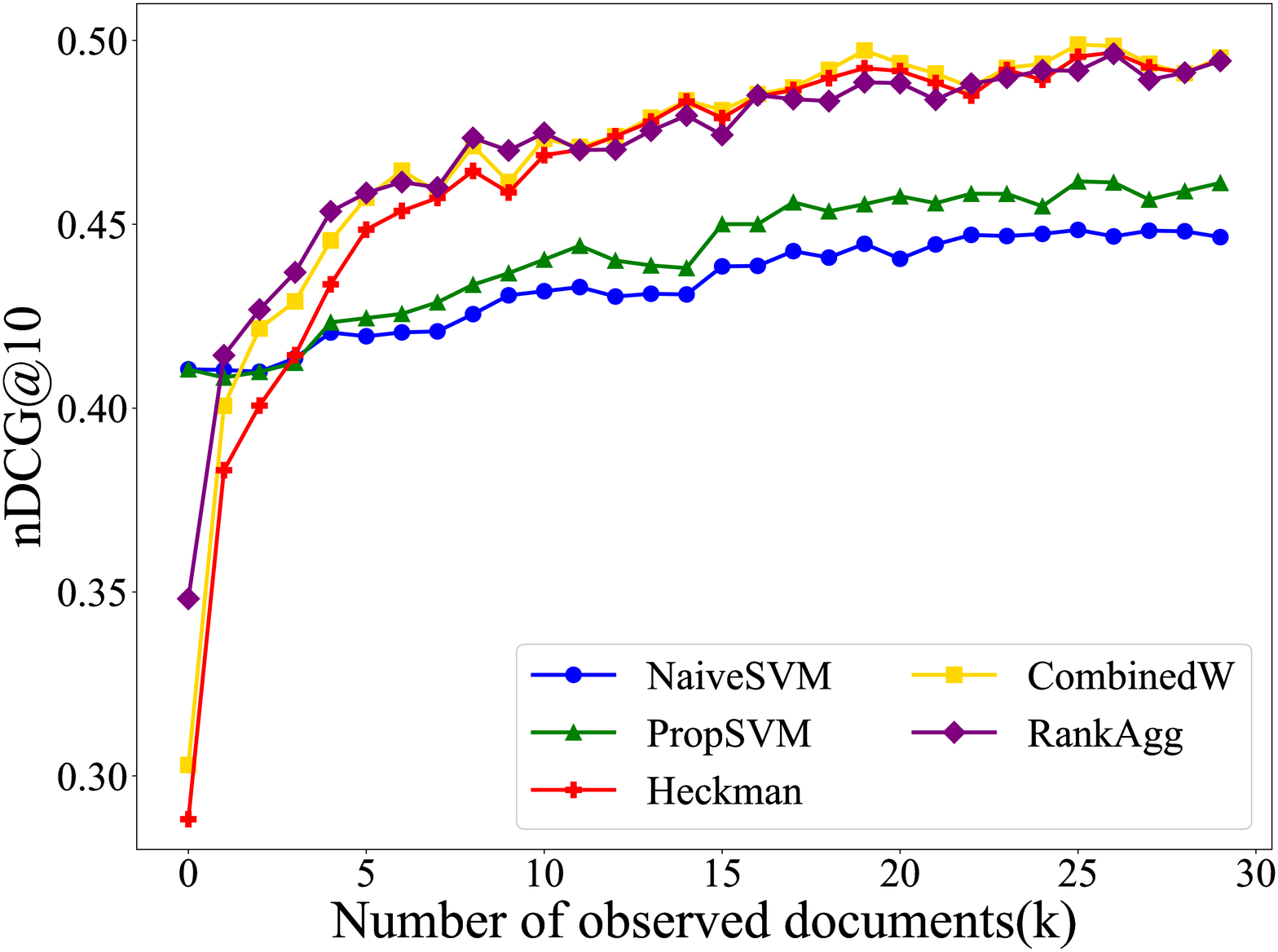}}\hfill
    \subfloat[nDCG@10, $\eta = 1$]{\label{sfig:bias_g}\includegraphics[width=.25\textwidth, height=0.21\textwidth]{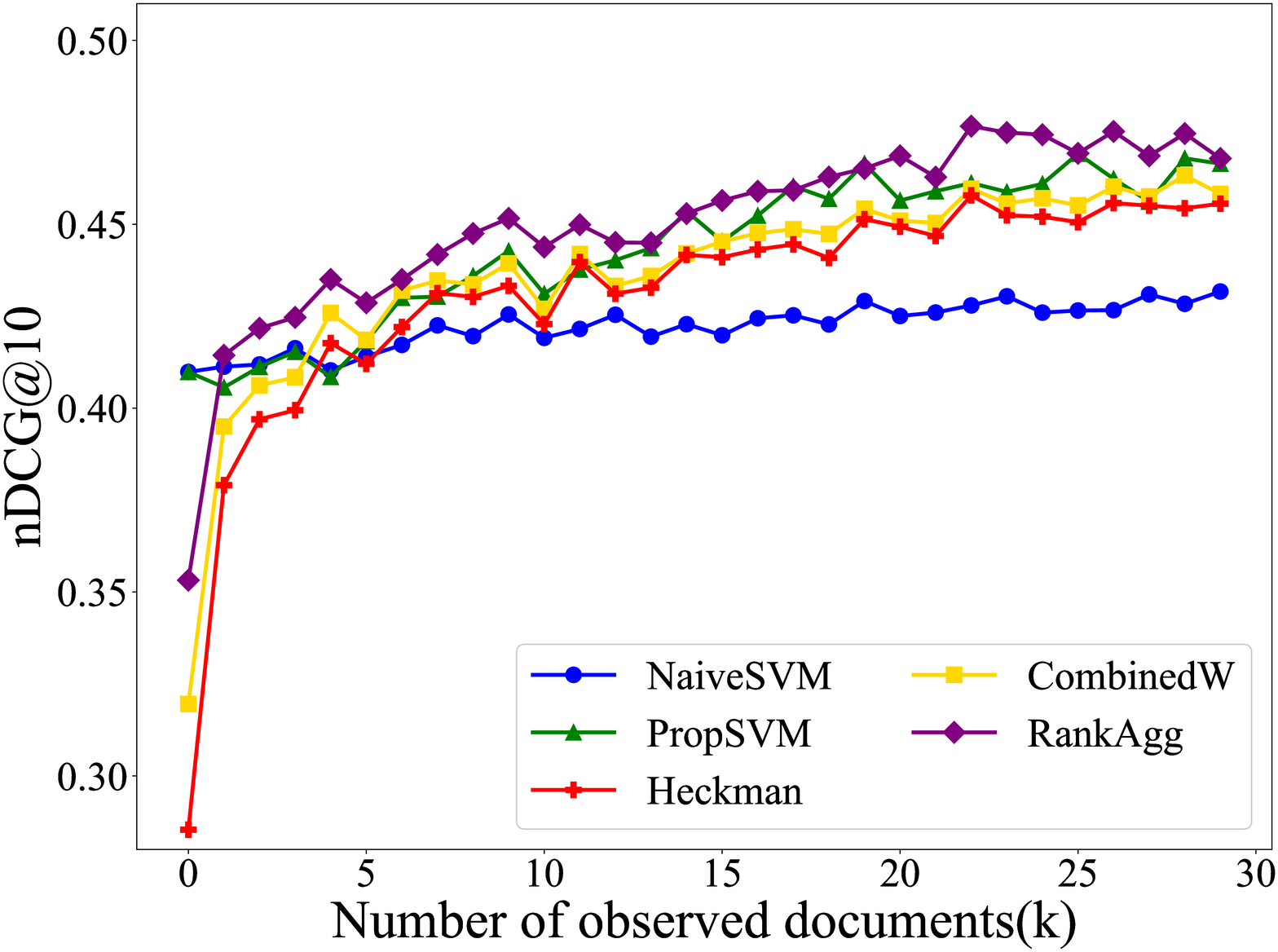}}\hfill
    \subfloat[nDCG@10, $\eta = 1.5$]{\label{sfig:bias_h}\includegraphics[width=.25\textwidth, height=0.21\textwidth]{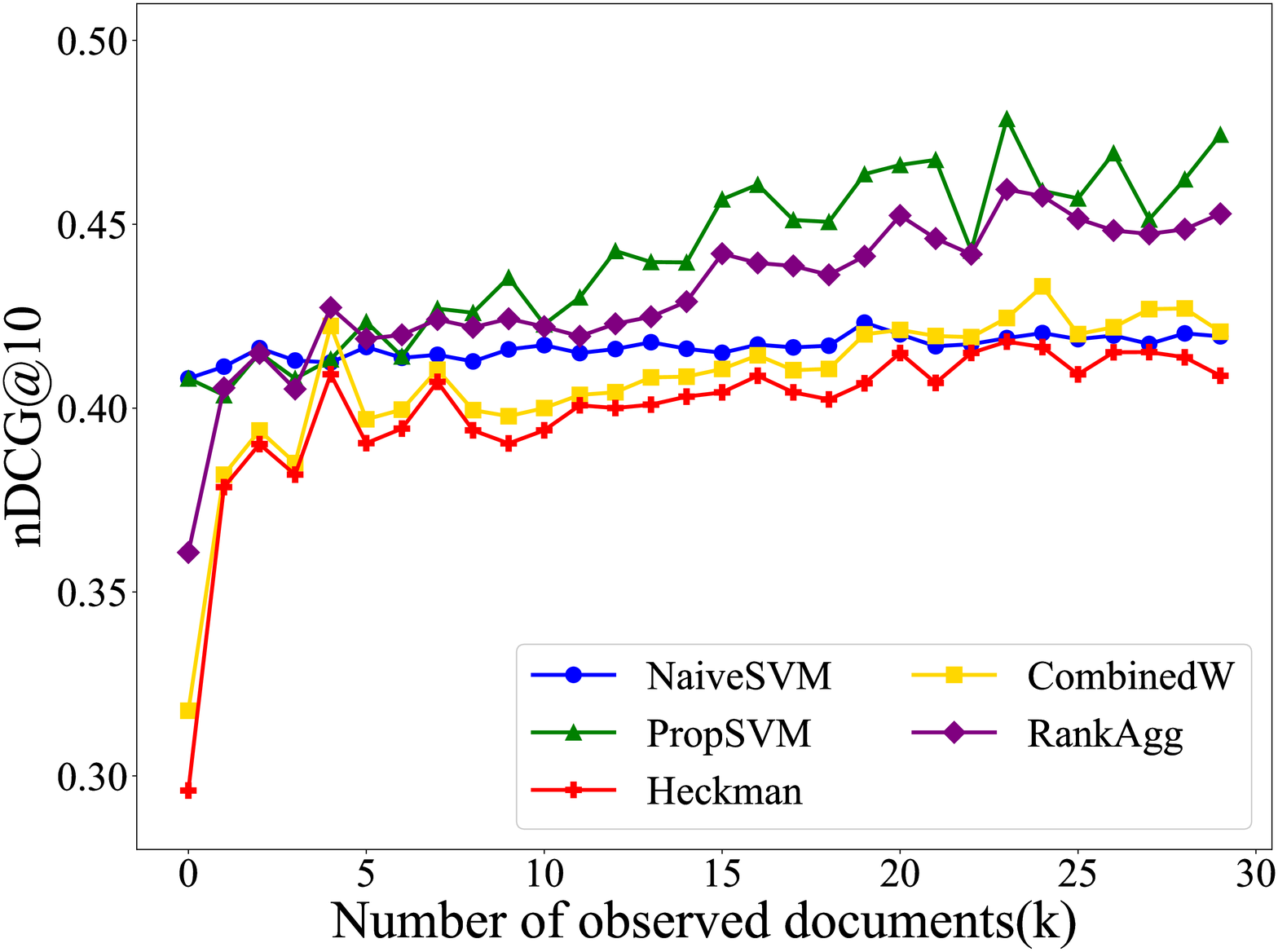}}\\
    \vspace{-0.2cm}
    \caption{The performance of LTR algorithms on set 2. Lower is better for ARRR, higher is better for nDCG@10.}
    \label{fig:fig_bias}
\end{figure*}

When training \textit{Propensity $SVM^{rank}$}, we apply an Inverse Propensity Score for clicked documents
$Q(o(y)=1| \textbf{x}, \mathbf{\Bar{y}},r)= (\frac{1}{rank(y|\mathbf{\Bar{y}})})^{\eta}$
where $o$ and $r$ represent whether a document is observed and relevant respectively, following \citet{joachims-wsdm17}. $Q(o(y)=1| \textbf{x}, {\mathbf{\Bar{y}}},r)$ is the propensity score denoting the marginal probability of observing the relevance of result $y$ for query $\mathbf{x}$ if the user was presented the ranking ${\mathbf{\Bar{y}}}$, and $\eta$ indicates the severity of position bias. 

\textit{$ Heckman^{rank}$} 
is implemented following the steps described in section \ref{sec:solution}. In step 1, the documents that appear among the $n$ shown results for each query are considered observed
($o_{x,y} = 1$), and the 
remainder as not-observed
($o_{x,y} = 0$). It is important to note that other LTR algorithms throw away the documents with $o_{x,y} = 0$ in training, while we do not. 
In our implementation $\bm{Z}$ only includes the feature set common to $\bm{F}_{x,y}$. 
For the ensemble methods, the selection bias recovery algorithm $A_s$ is \textit{$ Heckman^{rank}$} and the position bias recovery algorithm $A_p$ is \textit{Propensity $SVM^{rank}$}.

Given the model learned during training, each algorithm ranks the documents in the test set. In the following subsections, we evaluate each algorithm performance under different scenarios. For evaluation, the (noiseless) clicked documents in the test set are considered to be relevant documents, and the average rank of relevant results (ARRR) across queries along with nDCG@10 is our metric to evaluate each algorithm's performance.

\subsection{Experimental results on set 2}
\label{sec:Experimental results}

Here, we evaluate the performance of each algorithm under different levels of position bias ($\eta=0, 0.5, 1, 1.5, 2$) and when clicks are \textit{noisy} or \textit{noiseless} ($0\%$, $10\%$, $20\%$ and $30\%$ noise). In each case, we use ARRR and nDCG@10. %

\subsubsection{\textbf{Effect of position bias}}
\label{subsec.1}

Figure \ref{fig:fig_bias} illustrates the performance of all LTR algorithms and ensembles for varying degrees of position bias $(\eta\in \{0, 0.5, 1, 1.5\})$. %
Figures \ref{sfig:bias_a}, \ref{sfig:bias_b}, \ref{sfig:bias_c} and \ref{sfig:bias_d} show the performance as ARRR. Figures \ref{sfig:bias_e}, \ref{sfig:bias_f}, \ref{sfig:bias_g} and \ref{sfig:bias_h} show nDCG@10. Due to space, we omit the figures for $\eta=2$ since $\eta=1.5$ captures the trend of $propensitySVM$ starting to work better than the other methods. $Heckman^{rank}$ suffers when there is severe position bias. %

Figures \ref{sfig:bias_a}-\ref{sfig:bias_d} illustrate that \textit{$Heckman^{rank}$} outperforms \textit{Propensity $SVM^{rank}$} in the absence of position bias ($\eta=0$), or when position bias is low ($\eta=0.5$) and moderate ($\eta=1$). The better performance of \textit{$Heckman^{rank}$} over \textit{Propensity $SVM^{rank}$} vanishes with increased position bias, such that at a high position bias level ($\eta=1.5$), \textit{$Heckman^{rank}$} falls behind \textit{Propensity $SVM^{rank}$}, but still outperforms {Naive $SVM^{rank}$}. The reason for this is that a high position bias results in a high click frequency for top-ranked documents, leaving low-ranked documents with a very small chance of being clicked. \textit{$Heckman^{rank}$} is designed to control for the probability of a document being observed. If top-ranked documents have a disproportionately higher density in click data relative to low-ranked documents, then the predicted probabilities in \textit{$Heckman^{rank}$} will also reflect this imbalance.
In terms of algorithms that address both position bias and selection bias, Figures \ref{sfig:bias_a}, \ref{sfig:bias_b} show that for $\eta=0, 0.5$, $combinedW$ and $RankAgg$ outperform both \textit{Propensity $SVM^{rank}$} and $Heckman^{rank}$ for $k \lessapprox 7$. Moreover, Figures \ref{sfig:bias_c} and \ref{sfig:bias_d} show that for $\eta=1$ and $\eta=1.5$ $RankAgg$ outperforms its component algorithms for almost all values of $k$.

\begin{figure*}[!h]
    \centering
    \subfloat[ARRR, $\eta = 0$]{\label{sfig:noisy_a}\includegraphics[width=.25\textwidth, height=0.21\textwidth]{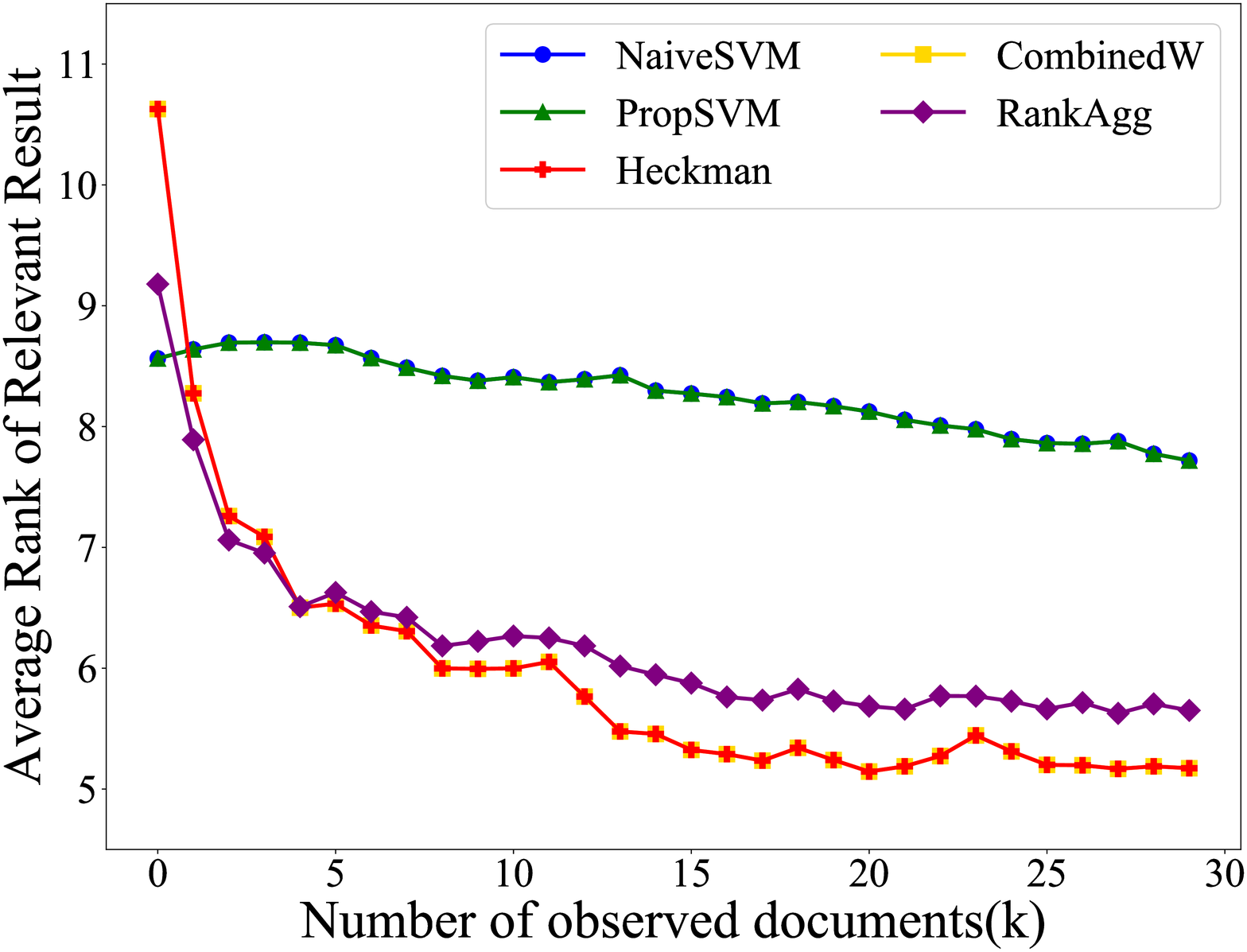}}\hfill
    \subfloat[ARRR, $\eta = 0.5$]{\label{sfig:noisy_b}\includegraphics[width=.25\textwidth, height=0.21\textwidth]{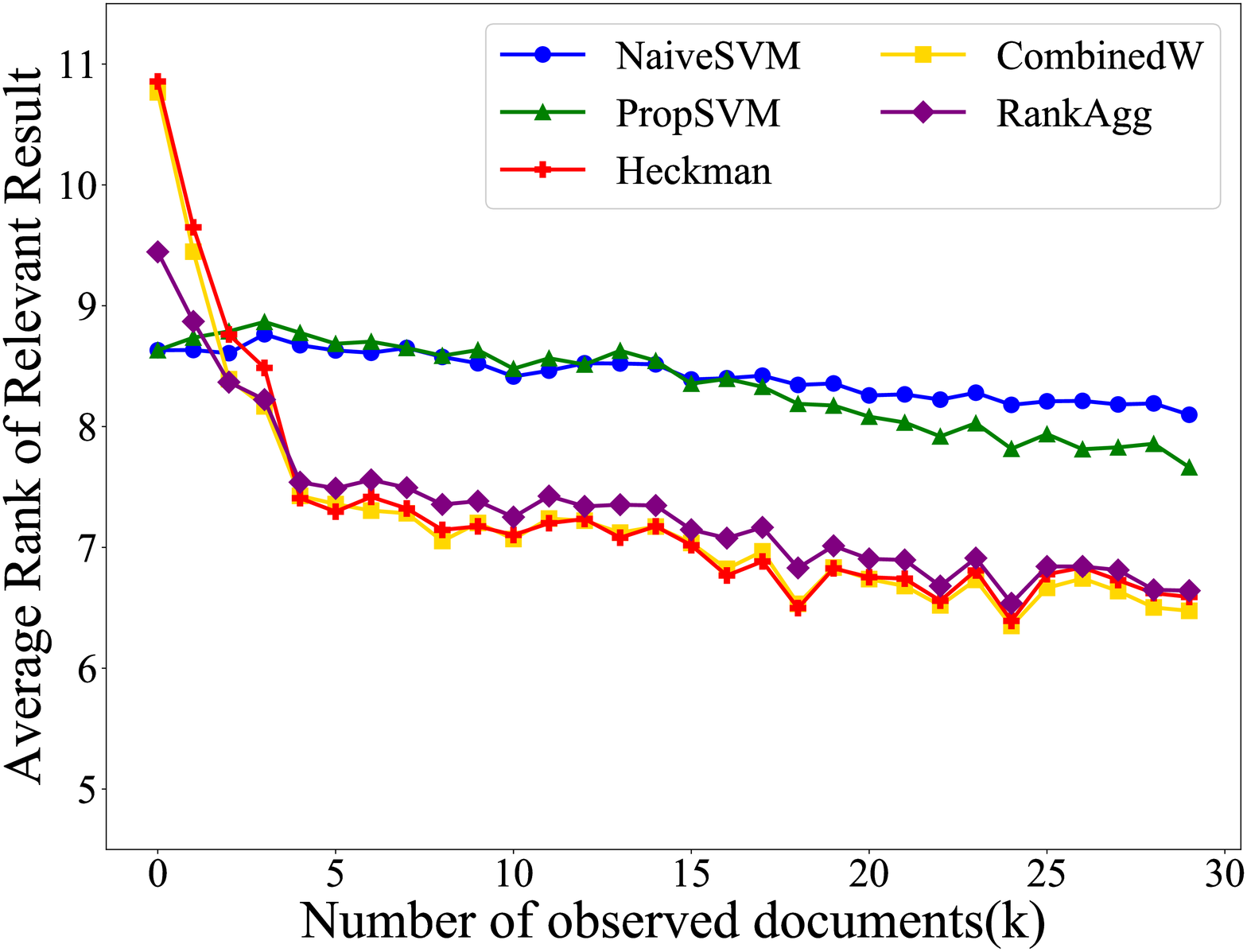}}\hfill
    \subfloat[ARRR, $\eta = 1$]{\label{sfig:noisy_c}\includegraphics[width=.25\textwidth, height=0.21\textwidth]{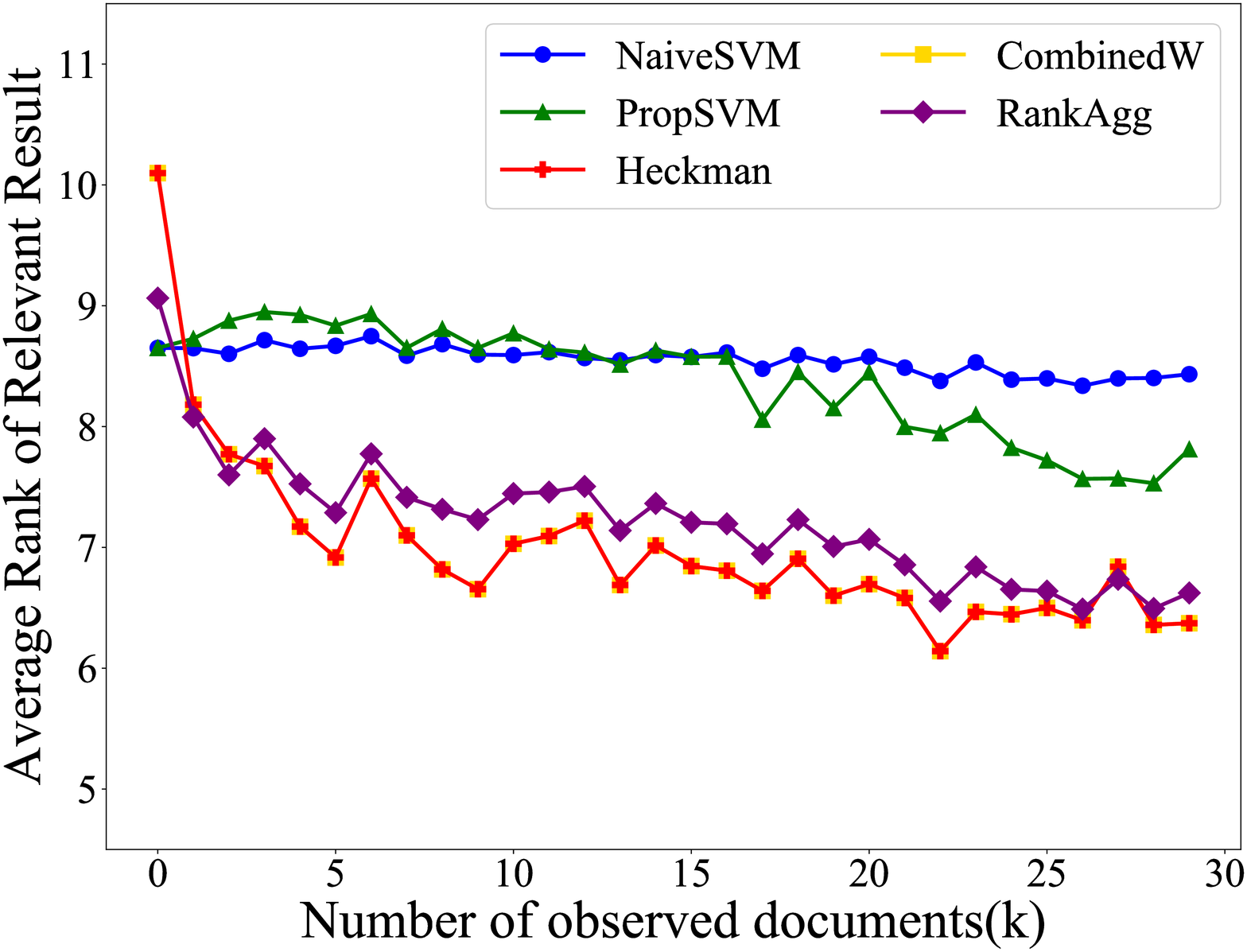}}\hfill
    \subfloat[ARRR, $\eta = 1.5$]{\label{sfig:noisy_d}\includegraphics[width=.25\textwidth, height=0.21\textwidth]{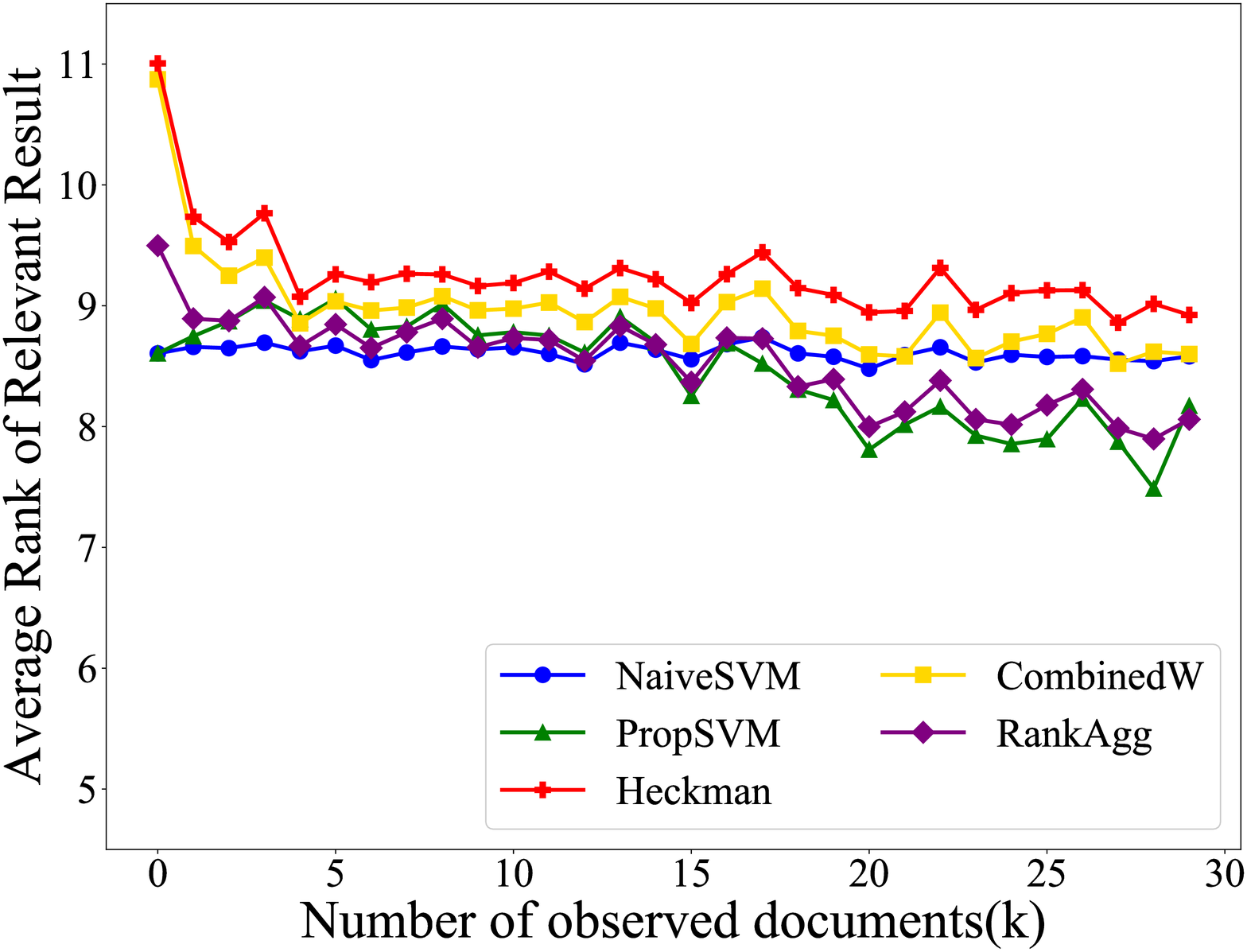}}\\
    
    \subfloat[nDCG@10, $\eta = 0$]{\label{sfig:noisy_e}\includegraphics[width=.25\textwidth, height=0.21\textwidth]{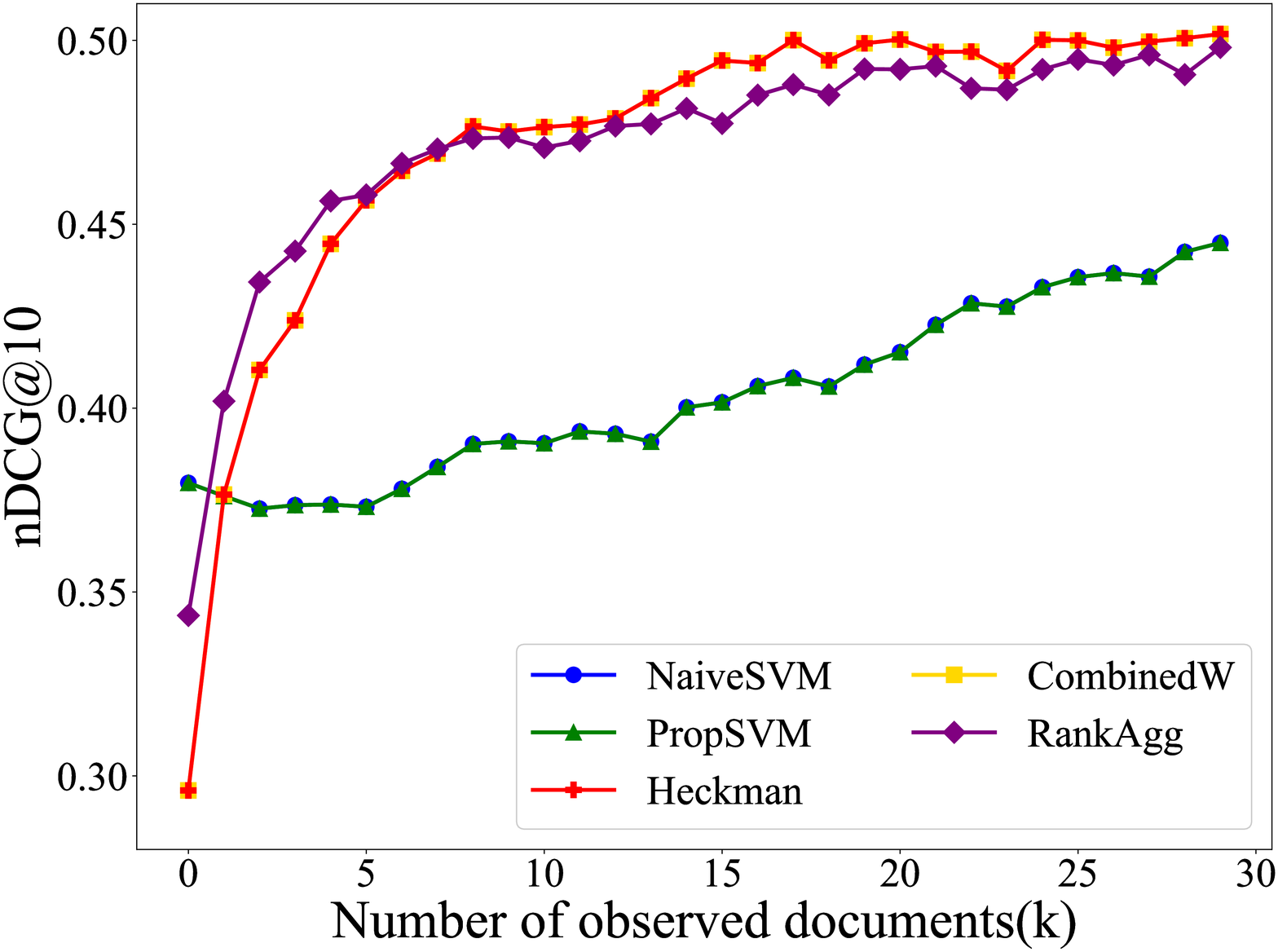}}\hfill
    \subfloat[nDCG@10, $\eta = 0.5$]{\label{sfig:noisy_f}\includegraphics[width=.25\textwidth, height=0.21\textwidth]{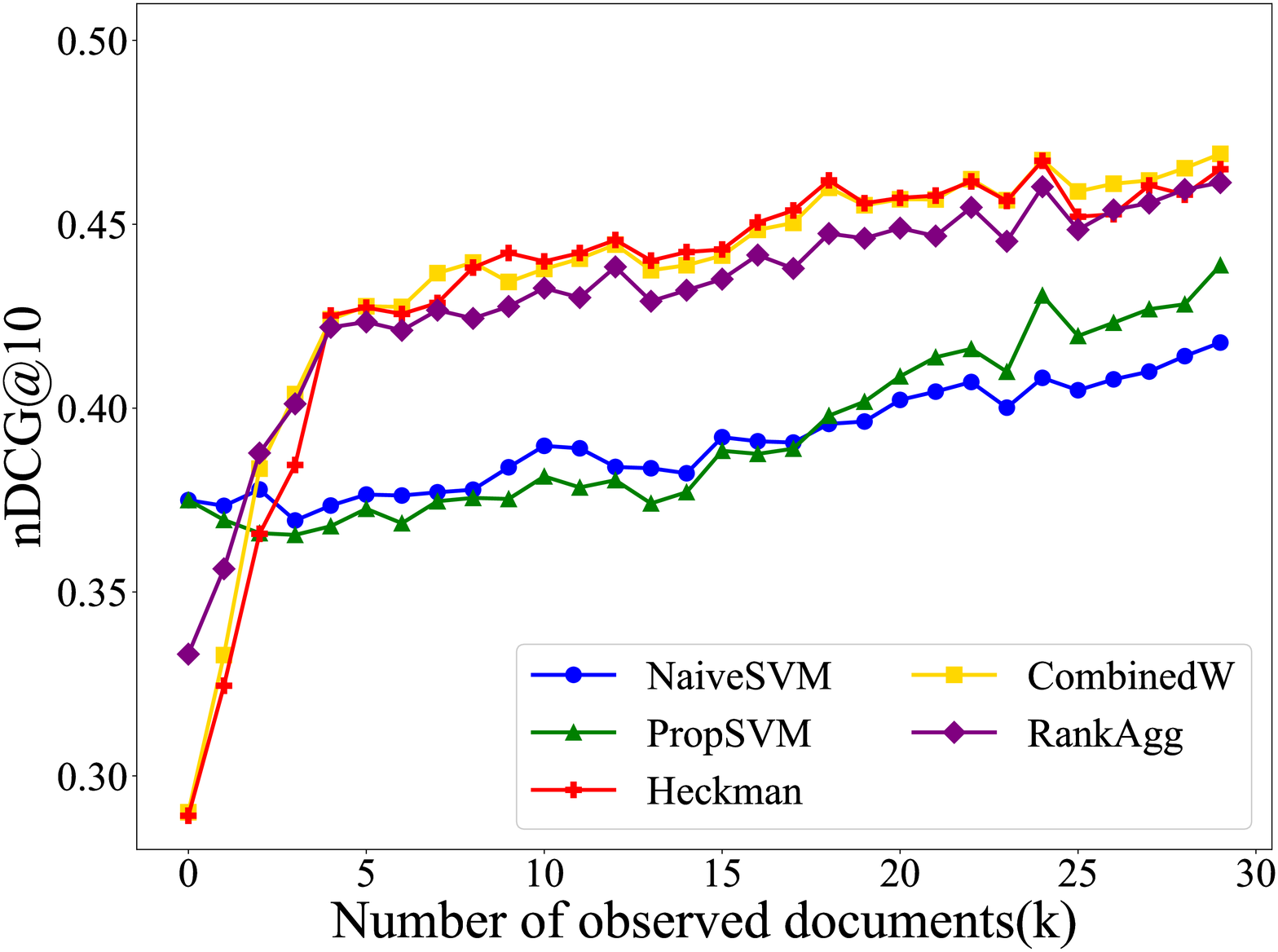}}\hfill
    \subfloat[nDCG@10, $\eta = 1$]{\label{sfig:noisy_g}\includegraphics[width=.25\textwidth, height=0.21\textwidth]{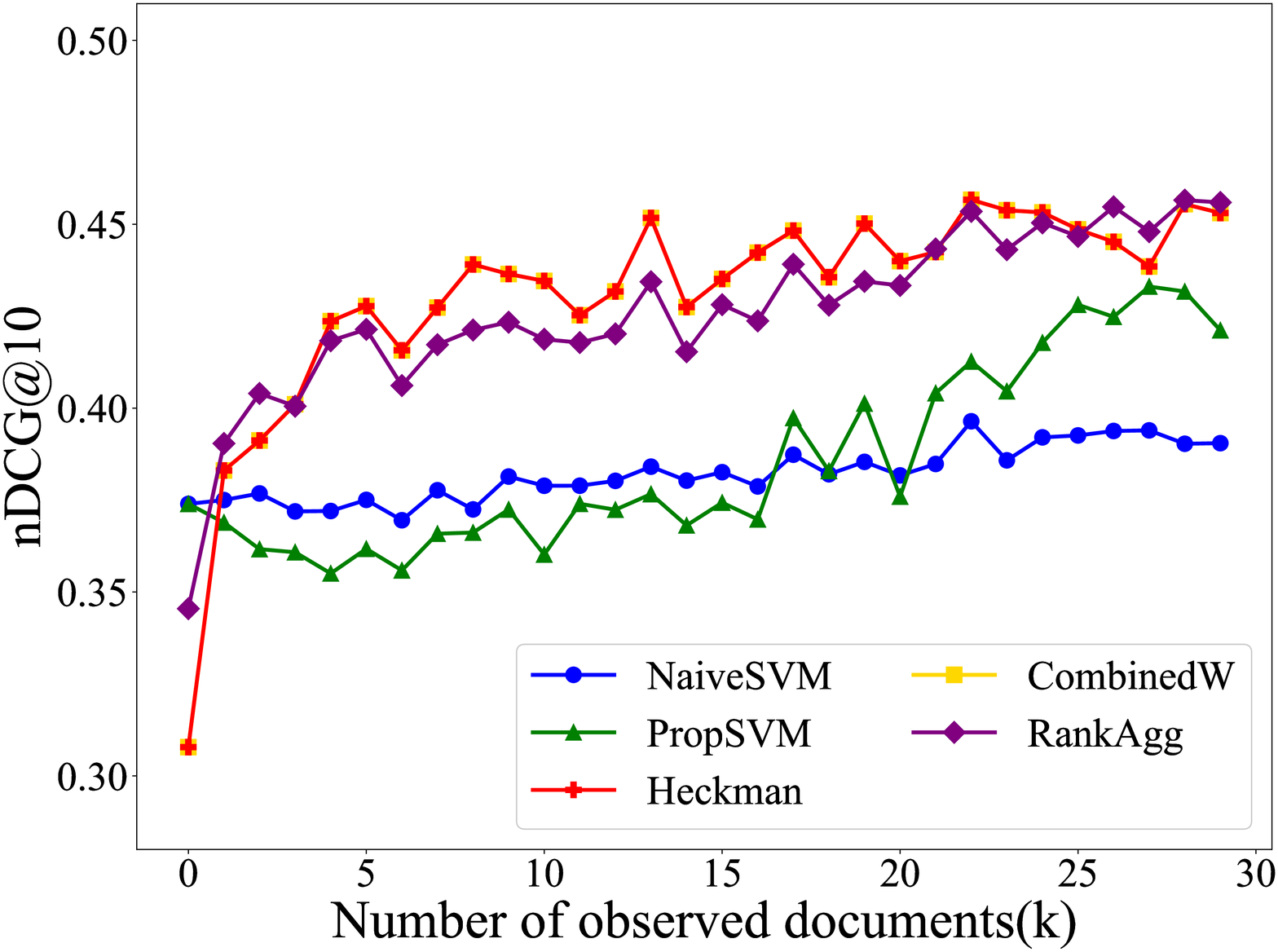}}\hfill
    \subfloat[nDCG@10, $\eta = 1.5$]{\label{sfig:noisy_h}\includegraphics[width=.25\textwidth, height=0.21\textwidth]{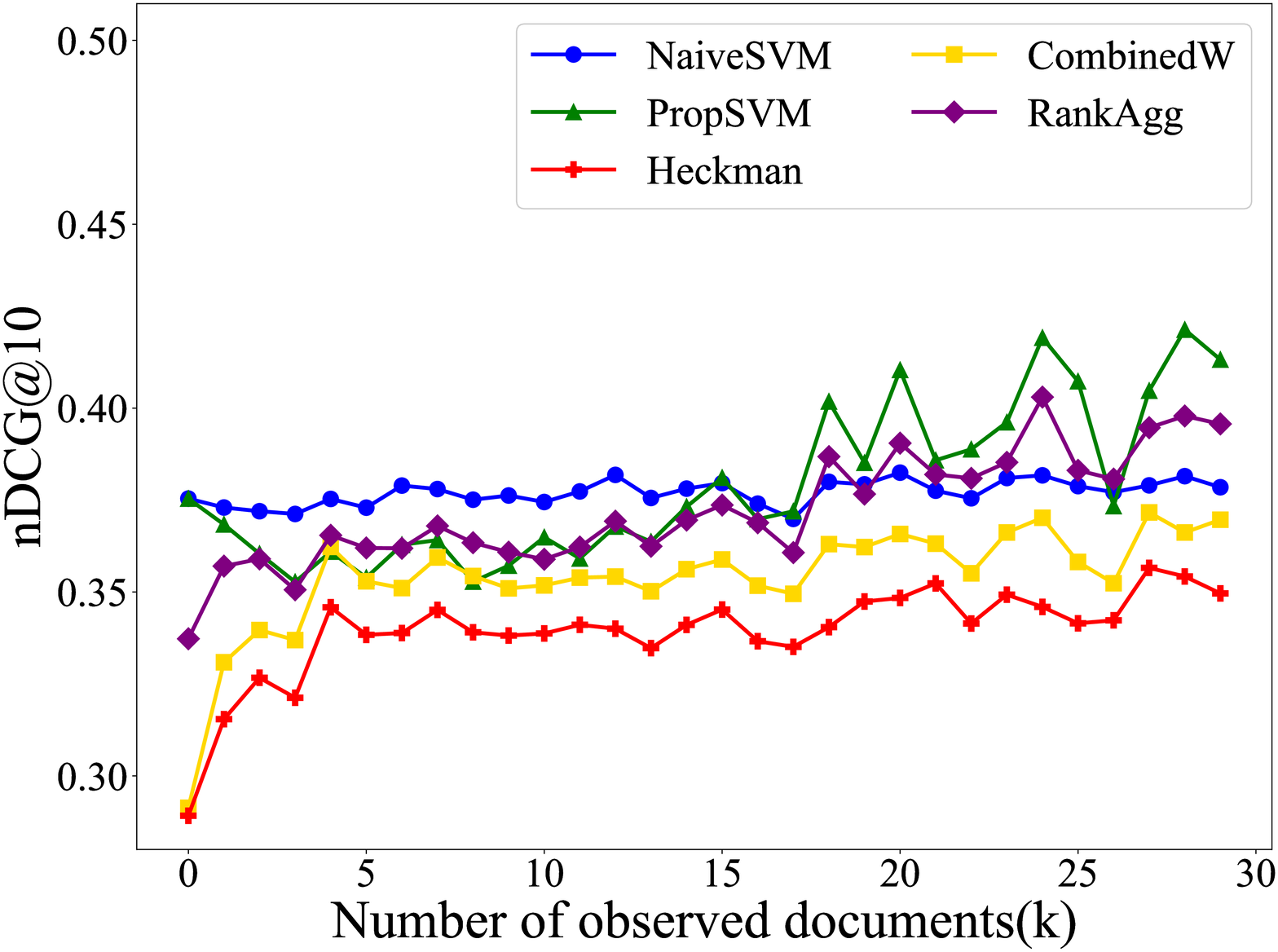}}\\
    \vspace{-0.3cm}
    \caption{The performance of LTR algorithms on set 2 under $10\%$ noisy clicks.}
    \label{fig:fig_noisy}
\end{figure*}

When $ARRR$ is the metric of interest, Figure \ref{sfig:bias_a}, \ref{sfig:bias_b}, \ref{sfig:bias_c} and \ref{sfig:bias_d} illustrate that \textit{$Heckman^{rank}$} outperforms \textit{Propensity $SVM^{rank}$} in the absence of position bias ($\eta=0$) and when position bias is low to moderate ($\eta=\{0.5, 1\}$), while it falls behind \textit{Propensity $SVM^{rank}$} when position bias increases ($\eta=1.5$). To compare it to the results for $nDCG@10$ illustrated in Figures \ref{sfig:bias_e}, \ref{sfig:bias_f}, \ref{sfig:bias_g} and \ref{sfig:bias_h}, \textit{$Heckman^{rank}$} appears to start lagging behind in performance at $\eta=1.5$.%
For the ensemble methods, Figures \ref{sfig:bias_e}, \ref{sfig:bias_f} illustrate that when $\eta=\{0, 0.5\}$ $combinedW$ and  $RankAgg$ outperform their  component algorithms for $k \lessapprox 10$. \ref{sfig:bias_g} demonstrates the better performance of $RankAgg$ to its component algorithms for all values of $k$ when $\eta=1$. However, for a severe position bias $\eta=1.5$, $combinedW$ and $RankAgg$ do not outperform their component algorithms for any value of $k$, but $RankAgg$ becomes the second best algorithm. Among the ensemble methods, $RankAgg$ is more robust to position bias than $combinedW$. %

Our main takeaways from this experiment are:  

\begin{itemize}
  \item Under small to no position bias ($\eta=0, 0.5$) \textit{$Heckman^{rank}$} outperforms \textit{Propensity $SVM^{rank}$} for both metrics. 
  \item Under moderate position bias ($\eta=1)$, while \textit{$Heckman^{rank}$} outperforms \textit{Propensity $SVM^{rank}$} for ARRR, it lags behind \textit{Propensity $SVM^{rank}$} for nDCG@10. 
  \item Under severe position bias ($\eta=1.5$), \textit{$Heckman^{rank}$} falls behind \textit{Propensity $SVM^{rank}$} for both $ARRR$ and $nDCG@10$.
  \item $RankAgg$ performs better than \textit{$Heckman^{rank}$} for all selection bias levels and it is more robust to position bias than $combinedW$. $combinedW$ surpasses \textit{$Heckman^{rank}$} under severe selection bias ($k \lessapprox 10$). 
\end{itemize}

\subsubsection{\textbf{Effect of click noise}}
\label{subs:noise}
Thus far, we have considered \textit{noiseless} clicks that are generated only over relevant documents. However, this is not a realistic assumption as users may also click on irrelevant documents. We now relax this assumption and allow for $10\%$ of the clicked documents to be irrelevant. %

When ARRR is the preferred metric, Figures \ref{sfig:noisy_a}, \ref{sfig:noisy_b}, \ref{sfig:noisy_c} and \ref{sfig:noisy_d} illustrate that \textit{$Heckman^{rank}$} outperforms \textit{Propensity $SVM^{rank}$} for $\eta=\{0, 0.5, 1\}$, while under higher position bias level ($\eta=1.5$), \textit{$Heckman^{rank}$} falls behind \textit{Propensity $SVM^{rank}$}. Comparing the noisy click performance to the noiseless one (Figures \ref{sfig:bias_a}, \ref{sfig:bias_b}, \ref{sfig:bias_c}), one can conclude that for $\eta=\{0, 0.5, 1\}$, \textit{Propensity $SVM^{rank}$} is highly affected by noise, while \textit{$Heckman^{rank}$} is much less affected. For example, Figure \ref{sfig:noisy_c} illustrates that for $\eta=1$, the better performance of \textit{$Heckman^{rank}$} over \textit{Propensity $SVM^{rank}$} is much more noticeable compared to \ref{sfig:bias_c} where clicks were noiseless. Interestingly, the ensembles $combinedW$ nor $RankAgg$, do not outperform the most successful algorithm in the presence of noisy clicks.

When nDCG@10 is the preferred metric, one can draw the same conclusions: while \textit{$Heckman^{rank}$} is more robust to noise and outperforms \textit{Propensity $SVM^{rank}$} for $\eta=\{0, 0.5, 1\}$, it fails to beat \textit{Propensity $SVM^{rank}$} for $\eta=1.5$. %
Another interesting point is that \textit{Propensity $SVM^{rank}$} is severely affected by noise when selection bias is high (low values of $k$), such that it even falls behind \textit{Naive $SVM^{rank}$}. This exemplifies how much selection bias can degrade the performance of LTR systems if they do not correct for it.

\begin{figure} [!h]
    \centering 
    \subfloat[Evaluation on ARRR.]{\label{sfig:c_noise_a}\includegraphics[width=.23\textwidth]{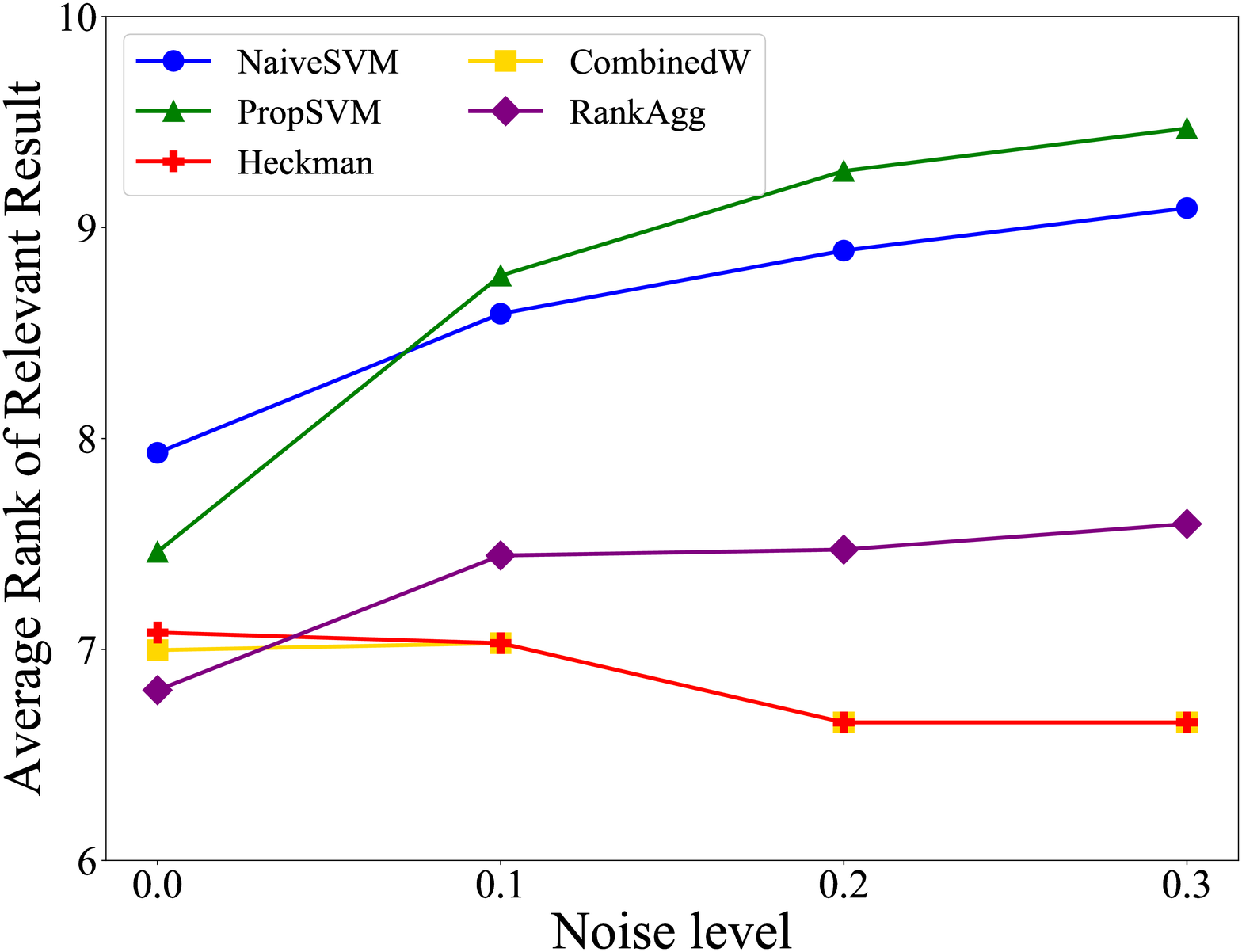}}\hfill
    \subfloat[Evaluation on nDCG@10.]{\label{sfig:c_noise_b}\includegraphics[width=.23\textwidth]{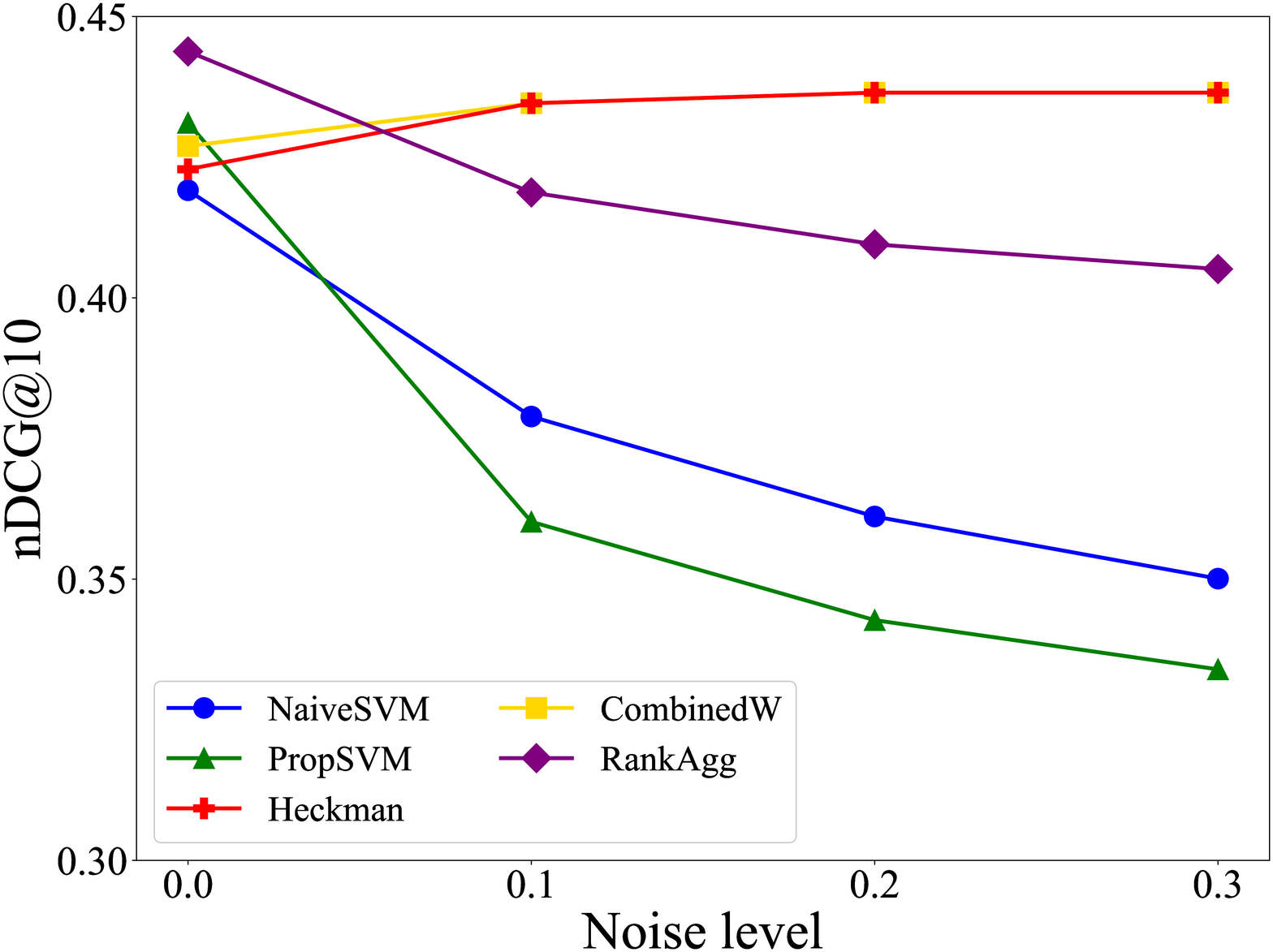}}\\
    \caption{Effect of noisy clicks for high selection bias ($k = 10$) and moderate position bias ($\eta = 1$).}
    \vspace{-0.3cm}
    \label{fig:c_noise}
\end{figure}

\begin{figure} [!h]
    \centering
    \subfloat[Evaluation on ARRR.]{\label{sfig:c_noise2_a}\includegraphics[width=.23\textwidth]{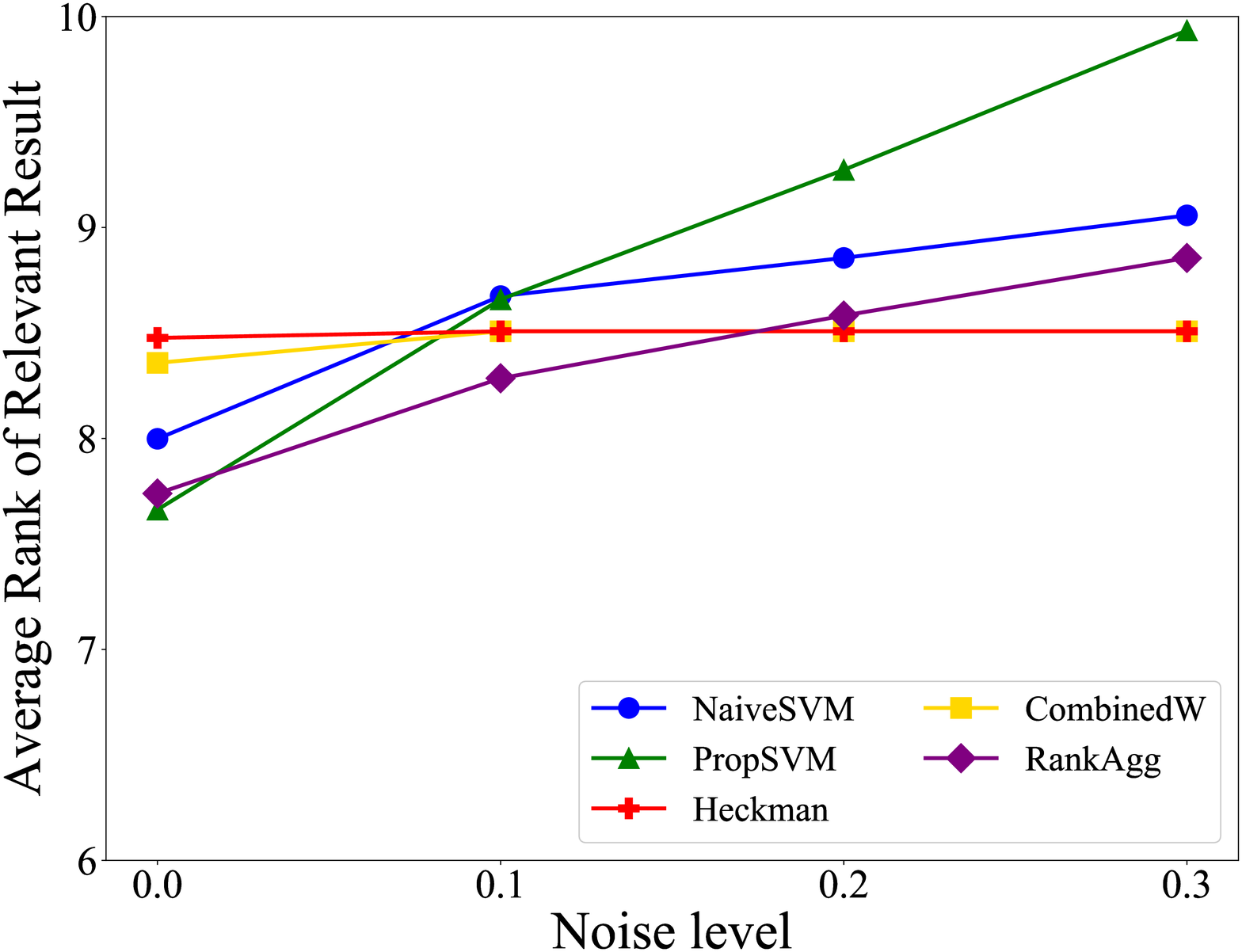}}\hfill
    \subfloat[Evaluation on nDCG@10.]{\label{sfig:c_noise2_b}\includegraphics[width=.23\textwidth]{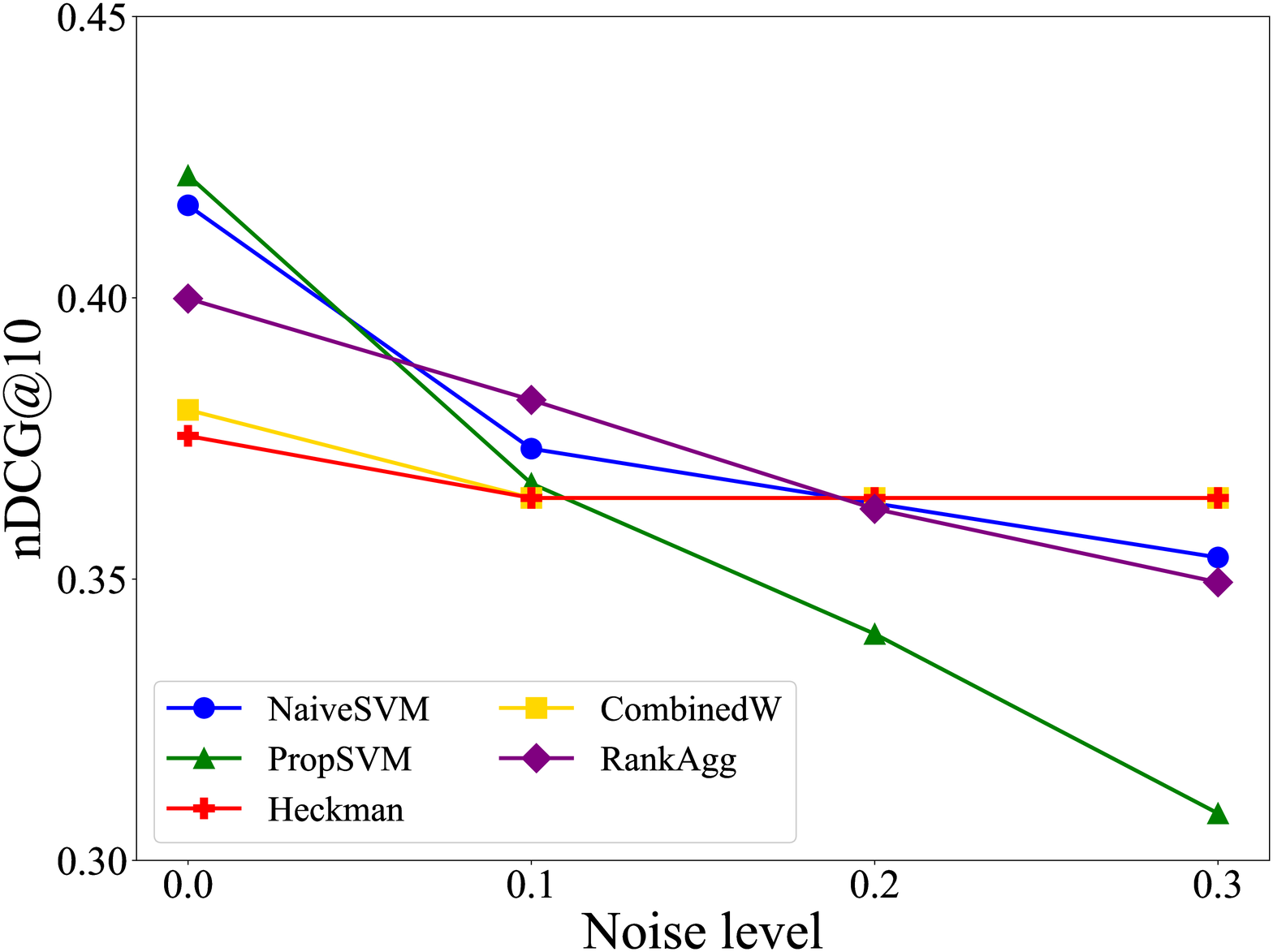}}\\
    \caption{Effect of noisy clicks for high selection bias ($k = 10$) and high position bias ($\eta = 2$).}
    \label{fig:c_noise2}
\end{figure}

\begin{figure*}
    \centering
    \subfloat[ARRR, $\eta = 0.5$]{\label{sfig:bias_a_set1}\includegraphics[width=.25\textwidth, height=0.21\textwidth]{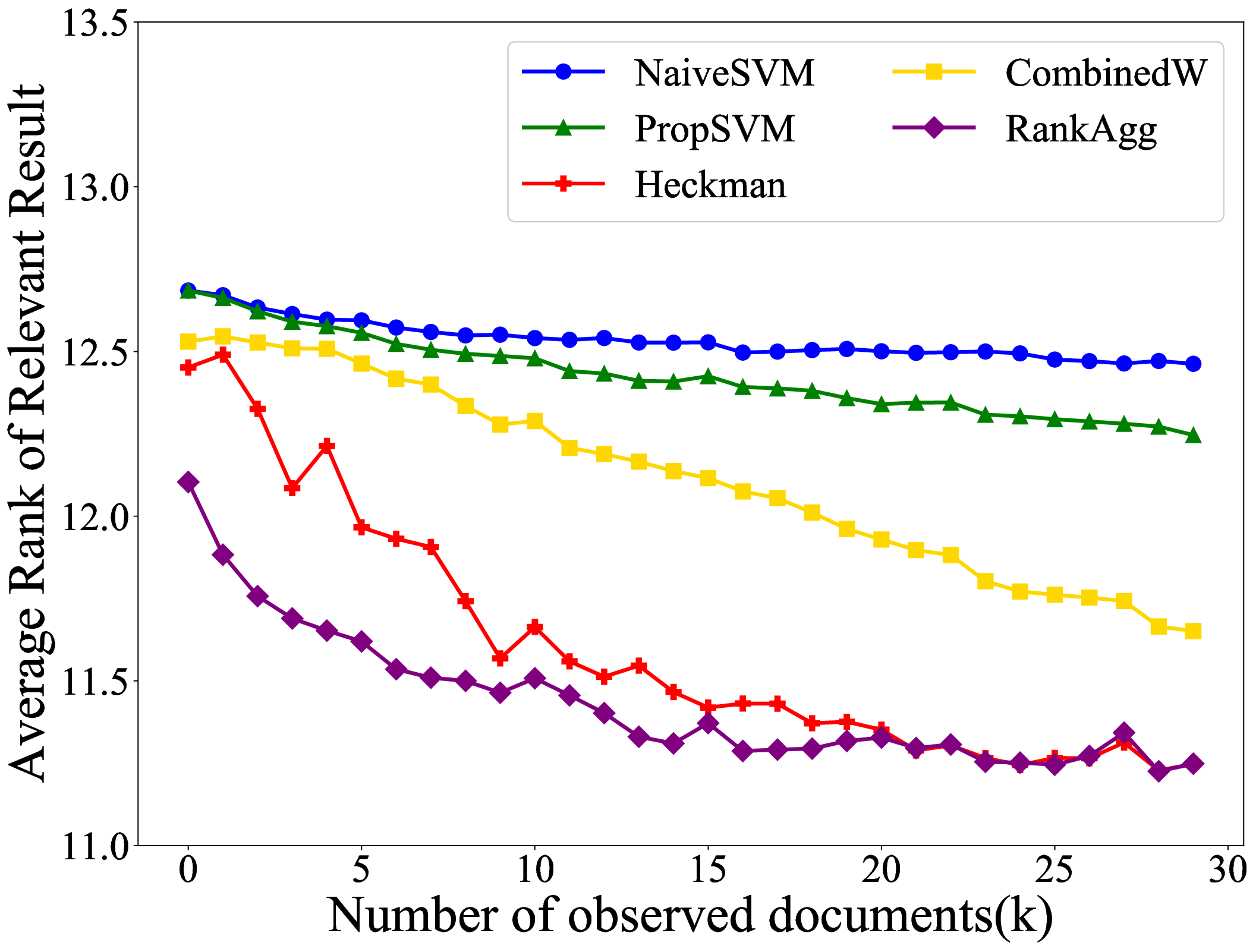}}\hfill
    \subfloat[ARRR, $\eta = 1$]{\label{sfig:bias_b_set1}\includegraphics[width=.25\textwidth, height=0.21\textwidth]{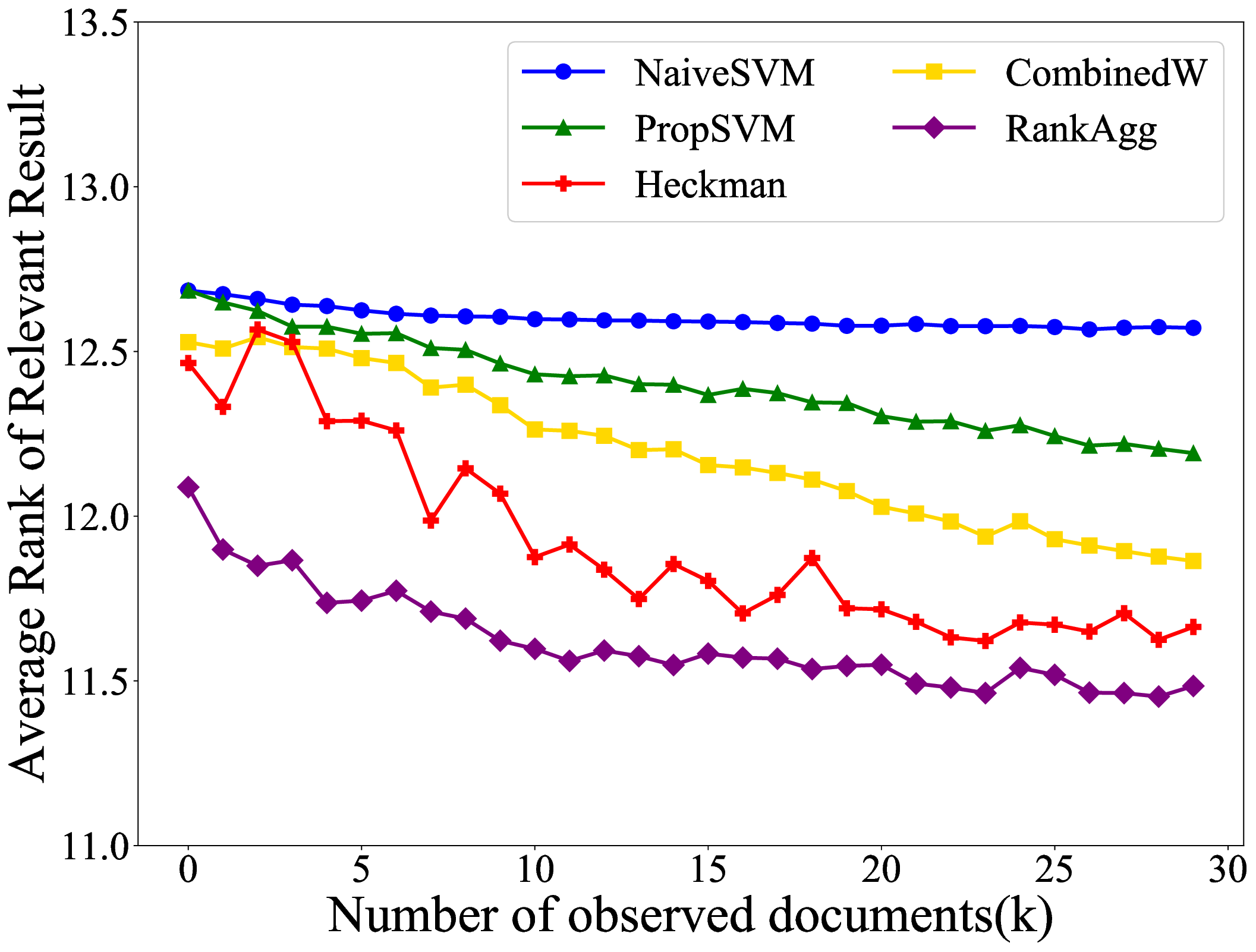}}\hfill
    \subfloat[ARRR, $\eta = 1.5$]{\label{sfig:bias_c_set1}\includegraphics[width=.25\textwidth, height=0.21\textwidth]{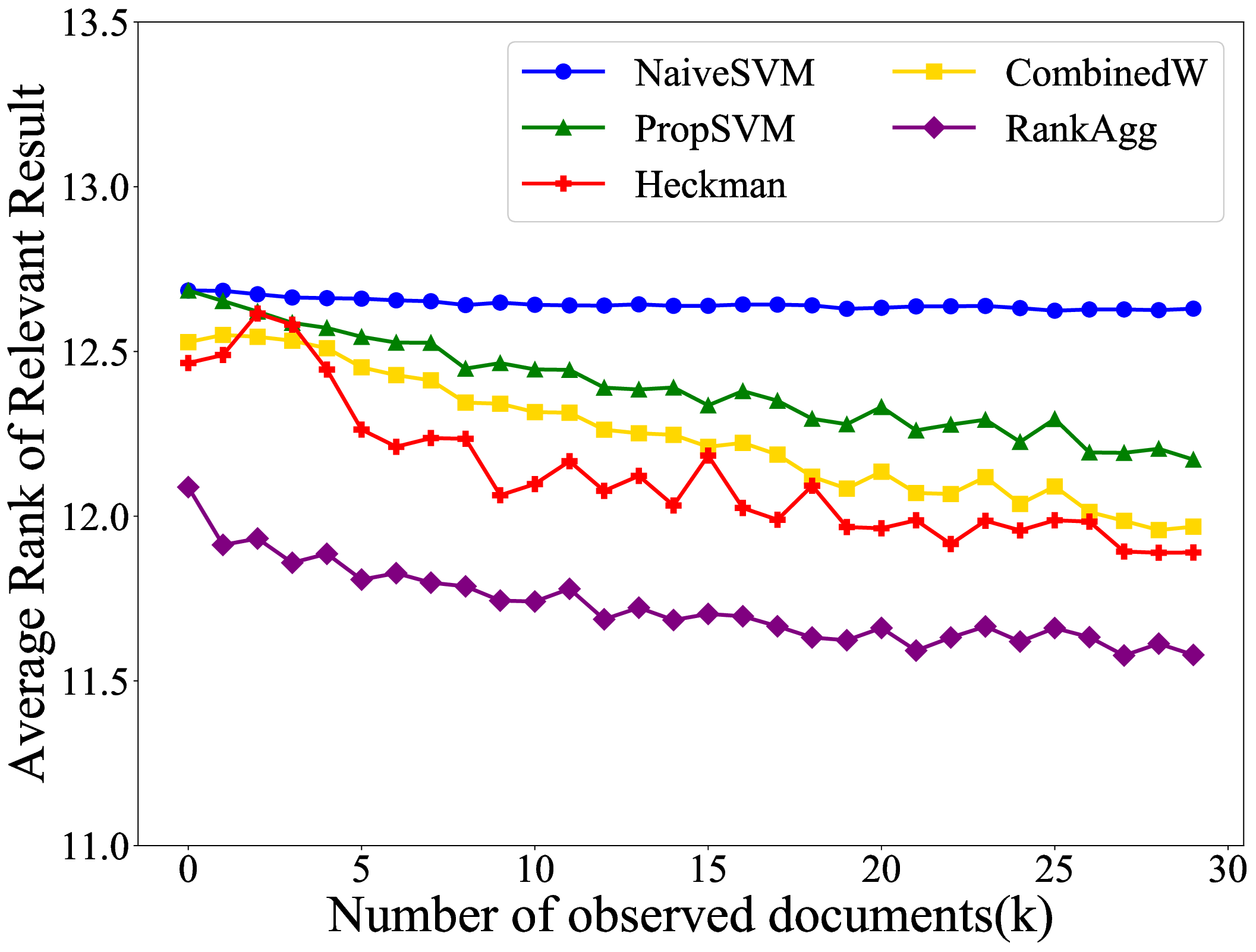}}\hfill
   \subfloat[ARRR, $\eta = 2$]{\label{sfig:bias_d_set1}\includegraphics[width=.25\textwidth, height=0.21\textwidth]{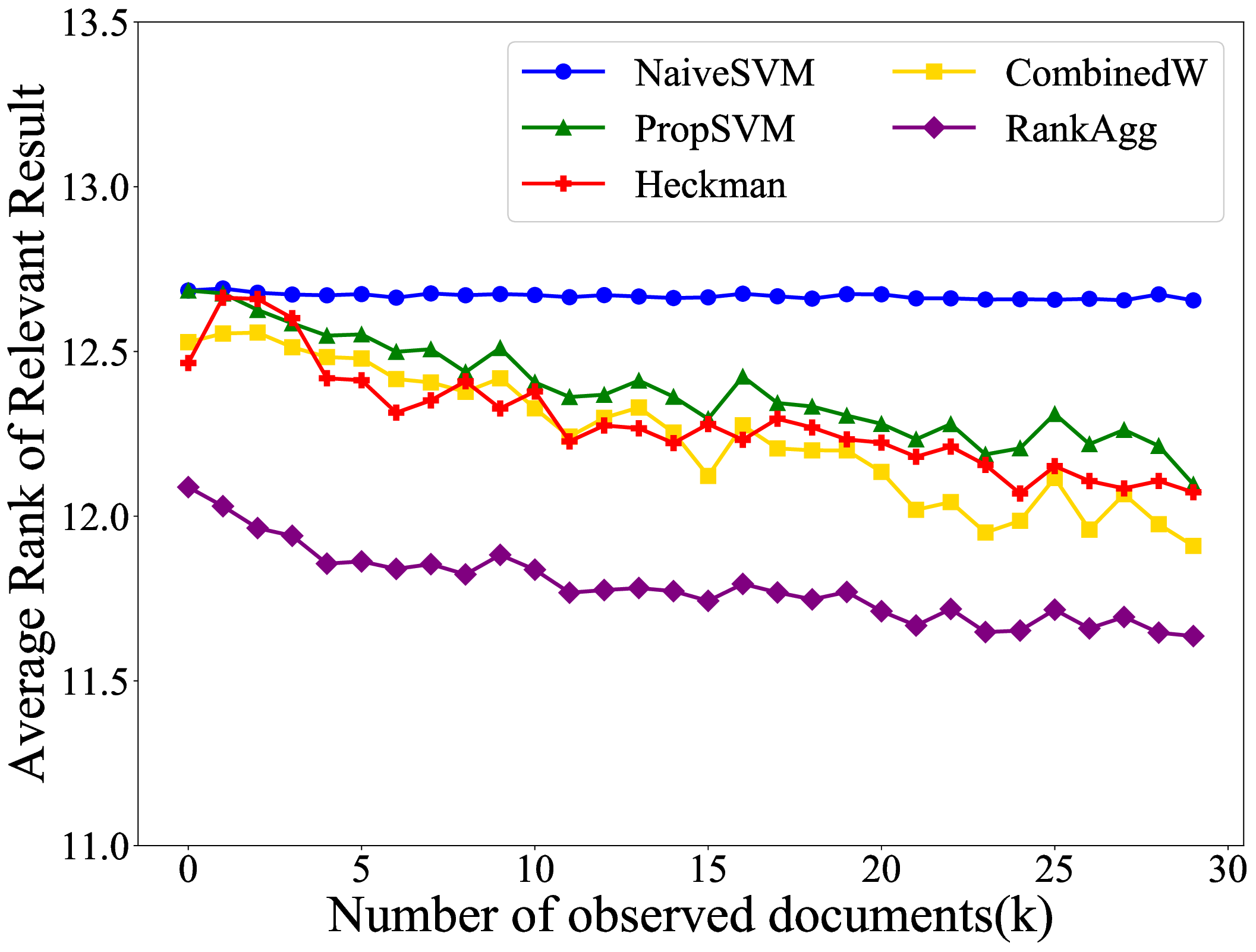}}\\
    \caption{The performance (ARRR) of LTR algorithms on set 1.}
    \label{fig:set1_fig_bias}
\end{figure*}

In the presence of $10\%$ noisy clicks, the main takeaways are: 

\begin{itemize}
  \item Under severe to moderate selection bias ($k \lessapprox 15$), \textit{Propensity $SVM^{rank}$} suffers a lot from the noise and it even falls behind \textit{Naive $SVM^{rank}$} for both $ARRR$ and $nDCG@10$.
  \item \textit{$Heckman^{rank}$} outperforms \textit{Propensity $SVM^{rank}$} when position bias is not severe ($\eta=\{0, 0.5, 1\}$) for both metrics. 
  \item Just like in the noiseless case, \textit{$Heckman^{rank}$} cannot surpass \textit{Propensity $SVM^{rank}$} under severe position bias ($\eta=1.5$). 
  \item $combinedW$ and $RankAgg$ surpass \textit{$Heckman^{rank}$} for a severe selection bias ($k \lessapprox 5$) when $\eta= \{0, 0.5\}$ for both $ARRR$ and $nDCG@10$. However, $RankAgg$ and $combinedW$ cannot beat \textit{Propensity $SVM^{rank}$} under high position bias.
\end{itemize} 

\subsubsection{\textbf{Effect of varying noise for $\eta=1$ and $\eta=2$}}
\label{subs:vary.noise}

In this section, we investigate whether our proposed models are robust to noise. Toward this goal, we varied the noise level from $0\%$ to $30\%$. Figures \ref{sfig:c_noise_a} and \ref{sfig:c_noise_b} show the performance of the LTR algorithms for different levels of noise, where $k=10$ and $\eta=1$. Under increasing noise, the performance of \textit{$Heckman^{rank}$} is relatively stable and even improves, while the performance of all other LTR algorithms degrades. Even \textit{Naive $SVM^{rank}$} is more robust to noise compared to \textit{Propensity $SVM^{rank}$}, which is different from the results by \citet{joachims-wsdm17} where no selection bias was considered. 
The reason could be that their evaluation %
is based on the assumption that all documents have a non-zero probability of being observed, while Figure \ref{sfig:c_noise_a} and \ref{sfig:c_noise_b} are under the condition that documents ranked bellow a certain cut-off ($k=10$) have a zero probability of being observed.

We also investigate the performance of LTR algorithms with respect to noise, when position bias is severe ($\eta=2$). As shown in Figure \ref{fig:c_noise2}, irrespective of metric of interest, \textit{$Heckman^{rank}$} is robust to varying noise, while the performance of all other algorithms degrades when the noise level increases. \textit{Propensity $SVM^{rank}$} falls behind all other algorithms in high level of noise. %
This implies that even though \textit{$Heckman^{rank}$} cannot surpass \textit{Propensity $SVM^{rank}$} when position bias is severe ($\eta=1.5, 2$) in noiseless environments, it clearly outperforms \textit{Propensity $SVM^{rank}$} in the presence of selection bias with noise.
This is an extremely useful property since in real world applications we cannot assume a noiseless environment.

\subsection{Experimental results on set 1}
\label{sec:Experimental results large}
To confirm the performance of our proposed methods on the larger set 1 with out-of-sample test data, %
we ran experiments varying position bias ($\eta =\{0.5, 1, 1.5, 2\}$) under noiseless clicks. %
The results on this dataset were even more promising, especially for high position bias. Figure~\ref{fig:set1_fig_bias} illustrates the ARRR performance of all algorithms. $Heckman^{rank}$ outperforms $Propensity SVM^{rank}$ for all position bias levels, though its strong performance decreases with increasing $\eta$. This is unlike set 2 where $Heckman^{rank}$ did not outperform $PropensitySVM^{rank}$ under high position bias.  
The ensemble $RankAgg$ outperforms both $Heckman^{rank}$ and $PropensitySVM^{rank}$ for all position and selection bias levels, while $combinedW$ outperforms $Propensity SVM^{rank}$ but does not surpass $Heckman^{rank}$. Moreover, the stronger performance of $Heckman^{rank}$ and $RankAgg$ over $PropensitySVM^{rank}$ is much more pronounced compared to set 2.

\section{Conclusion}

In this work, we formalized the problem of selection bias in learning-to-rank systems and proposed $Heckman^{rank}$ as an approach for correcting for selection bias. We also presented two ensemble methods that correct for both selection and position bias by combining the rankings of \textit{$Heckman^{rank}$} and \textit{Propensity $SVM^{rank}$}. 
Our extensive experiments on semi-synthetic datasets show that selection bias affects the performance of LTR systems and that \textit{$Heckman^{rank}$} performs better than existing approaches that correct for position bias but that do not address selection bias. %
Nonetheless, this performance decreases as the position bias increases.
At the same time, \textit{$Heckman^{rank}$} is more robust to noisy clicks even with severe position bias, while \textit{Propensity $SVM^{rank}$} is adversely affected by noisy clicks in the presence of selection bias and even falls behind \textit{Naive $SVM^{rank}$}. %
The ensemble methods, $combinedW$ and $RankAgg$, outperform \textit{$Heckman^{rank}$} for severe selection bias and zero to small position bias. %

Our initial study of selection bias suggests a number of promising future avenues for research. 
For example, our initial work considers only linear models but a Heckman-based solution to selection bias can be adapted to non-linear algorithms as well, including extensions that consider bias correction mechanisms specific to each learning-to-rank algorithm. Our experiments suggest that studying correction methods that jointly account for position bias and selection bias can potentially address the limitations of methods that only account for one. Finally, even though we specifically studied selection bias in the context of learning-to-rank systems, we expect that our methodology will have broader applications beyond LTR systems. %

\section*{Acknowledgements}

We thank John Fike and Chris Kanich for providing invaluable server support. We acknowledge Amazon Web Services for their support through AWS Cloud Credits for Research. This material is based on research sponsored in part by the Defense Advanced Research Projects Agency (DARPA) under contract number \\HR00111990114 and the National Science
Foundation under Grant No. 1801644. The views and conclusions contained
herein are those of the authors and should not be interpreted
as necessarily representing the official policies, either expressed or implied, of DARPA or the U.S. Government. The U.S. Government is authorized to reproduce and
distribute reprints for governmental purposes notwithstanding any copyright annotation therein.

\bibliography{references}


\begin{thebibliography}{45}


\ifx \showCODEN    \undefined \def \showCODEN     #1{\unskip}     \fi
\ifx \showDOI      \undefined \def \showDOI       #1{#1}\fi
\ifx \showISBNx    \undefined \def \showISBNx     #1{\unskip}     \fi
\ifx \showISBNxiii \undefined \def \showISBNxiii  #1{\unskip}     \fi
\ifx \showISSN     \undefined \def \showISSN      #1{\unskip}     \fi
\ifx \showLCCN     \undefined \def \showLCCN      #1{\unskip}     \fi
\ifx \shownote     \undefined \def \shownote      #1{#1}          \fi
\ifx \showarticletitle \undefined \def \showarticletitle #1{#1}   \fi
\ifx \showURL      \undefined \def \showURL       {\relax}        \fi
\providecommand\bibfield[2]{#2}
\providecommand\bibinfo[2]{#2}
\providecommand\natexlab[1]{#1}
\providecommand\showeprint[2][]{arXiv:#2}

\bibitem[\protect\citeauthoryear{Agarwal, Takatsu, Zaitsev, and
  Joachims}{Agarwal et~al\mbox{.}}{2019}]%
        {agarwal2019general}
\bibfield{author}{\bibinfo{person}{Aman Agarwal}, \bibinfo{person}{Kenta
  Takatsu}, \bibinfo{person}{Ivan Zaitsev}, {and} \bibinfo{person}{Thorsten
  Joachims}.} \bibinfo{year}{2019}\natexlab{}.
\newblock \showarticletitle{A General Framework for Counterfactual
  Learning-to-Rank}. In \bibinfo{booktitle}{\emph{ACM Conference on Research
  and Development in Information Retrieval (SIGIR)}}.
\newblock


\bibitem[\protect\citeauthoryear{Ai, Bi, Luo, Guo, and Croft}{Ai
  et~al\mbox{.}}{2018}]%
        {ai-sigir18}
\bibfield{author}{\bibinfo{person}{Qingyao Ai}, \bibinfo{person}{Keping Bi},
  \bibinfo{person}{Cheng Luo}, \bibinfo{person}{Jiafeng Guo}, {and}
  \bibinfo{person}{W~Bruce Croft}.} \bibinfo{year}{2018}\natexlab{}.
\newblock \showarticletitle{Unbiased Learning to Rank with Unbiased Propensity
  Estimation}.
\newblock \bibinfo{journal}{\emph{SIGIR}} (\bibinfo{year}{2018}).
\newblock


\bibitem[\protect\citeauthoryear{Bareinboim and Pearl}{Bareinboim and
  Pearl}{2012}]%
        {bareinboim-aistats12}
\bibfield{author}{\bibinfo{person}{Elias Bareinboim} {and}
  \bibinfo{person}{Judea Pearl}.} \bibinfo{year}{2012}\natexlab{}.
\newblock \showarticletitle{Controlling selection bias in causal inference}. In
  \bibinfo{booktitle}{\emph{AISTATS}}. \bibinfo{pages}{100--108}.
\newblock


\bibitem[\protect\citeauthoryear{Bareinboim and Pearl}{Bareinboim and
  Pearl}{2016}]%
        {bareinboim-pnas16}
\bibfield{author}{\bibinfo{person}{Elias Bareinboim} {and}
  \bibinfo{person}{Judea Pearl}.} \bibinfo{year}{2016}\natexlab{}.
\newblock \showarticletitle{Causal inference and the data-fusion problem}.
\newblock \bibinfo{journal}{\emph{Proceedings of the National Academy of
  Sciences}} \bibinfo{volume}{113}, \bibinfo{number}{27}
  (\bibinfo{year}{2016}), \bibinfo{pages}{7345--7352}.
\newblock


\bibitem[\protect\citeauthoryear{Bareinboim and Tian}{Bareinboim and
  Tian}{2015}]%
        {bareinboim2015recovering}
\bibfield{author}{\bibinfo{person}{Elias Bareinboim} {and} \bibinfo{person}{Jin
  Tian}.} \bibinfo{year}{2015}\natexlab{}.
\newblock \showarticletitle{Recovering causal effects from selection bias}. In
  \bibinfo{booktitle}{\emph{Twenty-Ninth AAAI Conference on Artificial
  Intelligence}}.
\newblock


\bibitem[\protect\citeauthoryear{Bareinboim, Tian, and Pearl}{Bareinboim
  et~al\mbox{.}}{2014}]%
        {bareinboim-aaai14}
\bibfield{author}{\bibinfo{person}{Elias Bareinboim}, \bibinfo{person}{Jin
  Tian}, {and} \bibinfo{person}{Judea Pearl}.} \bibinfo{year}{2014}\natexlab{}.
\newblock \showarticletitle{Recovering from Selection Bias in Causal and
  Statistical Inference.}. In \bibinfo{booktitle}{\emph{AAAI}}.
  \bibinfo{pages}{2410--2416}.
\newblock


\bibitem[\protect\citeauthoryear{Borisov, Markov, de~Rijke, and
  Serdyukov}{Borisov et~al\mbox{.}}{2016}]%
        {borisov2016neural}
\bibfield{author}{\bibinfo{person}{Alexey Borisov}, \bibinfo{person}{Ilya
  Markov}, \bibinfo{person}{Maarten de Rijke}, {and} \bibinfo{person}{Pavel
  Serdyukov}.} \bibinfo{year}{2016}\natexlab{}.
\newblock \showarticletitle{A neural click model for web search}. In
  \bibinfo{booktitle}{\emph{Proceedings of the 25th International Conference on
  World Wide Web}}. International World Wide Web Conferences Steering
  Committee, \bibinfo{pages}{531--541}.
\newblock


\bibitem[\protect\citeauthoryear{Celma and Cano}{Celma and Cano}{2008}]%
        {celma2008hits}
\bibfield{author}{\bibinfo{person}{{\`O}scar Celma} {and}
  \bibinfo{person}{Pedro Cano}.} \bibinfo{year}{2008}\natexlab{}.
\newblock \showarticletitle{From hits to niches?: or how popular artists can
  bias music recommendation and discovery}. In
  \bibinfo{booktitle}{\emph{Proceedings of the 2nd KDD Workshop on Large-Scale
  Recommender Systems and the Netflix Prize Competition}}. ACM,
  \bibinfo{pages}{5}.
\newblock


\bibitem[\protect\citeauthoryear{Chaney, Stewart, and Engelhardt}{Chaney
  et~al\mbox{.}}{2018}]%
        {chaney-recsys18}
\bibfield{author}{\bibinfo{person}{Allison~JB Chaney},
  \bibinfo{person}{Brandon~M Stewart}, {and} \bibinfo{person}{Barbara~E
  Engelhardt}.} \bibinfo{year}{2018}\natexlab{}.
\newblock \showarticletitle{How Algorithmic Confounding in Recommendation
  Systems Increases Homogeneity and Decreases Utility}.
\newblock \bibinfo{journal}{\emph{RecSys}} (\bibinfo{year}{2018}).
\newblock


\bibitem[\protect\citeauthoryear{Chapelle and Chang}{Chapelle and
  Chang}{2011}]%
        {chapelle-11yahoo}
\bibfield{author}{\bibinfo{person}{Olivier Chapelle} {and} \bibinfo{person}{Yi
  Chang}.} \bibinfo{year}{2011}\natexlab{}.
\newblock \showarticletitle{Yahoo! learning to rank challenge overview}. In
  \bibinfo{booktitle}{\emph{Proceedings of the Learning to Rank Challenge}}.
  \bibinfo{pages}{1--24}.
\newblock


\bibitem[\protect\citeauthoryear{Chapelle, Joachims, Radlinski, and
  Yue}{Chapelle et~al\mbox{.}}{2012}]%
        {chapelle2012large}
\bibfield{author}{\bibinfo{person}{Olivier Chapelle}, \bibinfo{person}{Thorsten
  Joachims}, \bibinfo{person}{Filip Radlinski}, {and} \bibinfo{person}{Yisong
  Yue}.} \bibinfo{year}{2012}\natexlab{}.
\newblock \showarticletitle{Large-scale validation and analysis of interleaved
  search evaluation}.
\newblock \bibinfo{journal}{\emph{ACM Transactions on Information Systems
  (TOIS)}} \bibinfo{volume}{30}, \bibinfo{number}{1} (\bibinfo{year}{2012}),
  \bibinfo{pages}{6}.
\newblock


\bibitem[\protect\citeauthoryear{Chapelle and Zhang}{Chapelle and
  Zhang}{2009}]%
        {chapelle2009dynamic}
\bibfield{author}{\bibinfo{person}{Olivier Chapelle} {and} \bibinfo{person}{Ya
  Zhang}.} \bibinfo{year}{2009}\natexlab{}.
\newblock \showarticletitle{A dynamic bayesian network click model for web
  search ranking}. In \bibinfo{booktitle}{\emph{Proceedings of the 18th
  international conference on World wide web}}. ACM, \bibinfo{pages}{1--10}.
\newblock


\bibitem[\protect\citeauthoryear{Chen, De~Gemmis, Felfernig, Lops, Ricci, and
  Semeraro}{Chen et~al\mbox{.}}{2013}]%
        {chen2013human}
\bibfield{author}{\bibinfo{person}{Li Chen}, \bibinfo{person}{Marco De~Gemmis},
  \bibinfo{person}{Alexander Felfernig}, \bibinfo{person}{Pasquale Lops},
  \bibinfo{person}{Francesco Ricci}, {and} \bibinfo{person}{Giovanni
  Semeraro}.} \bibinfo{year}{2013}\natexlab{}.
\newblock \showarticletitle{Human decision making and recommender systems}.
\newblock \bibinfo{journal}{\emph{ACM Transactions on Interactive Intelligent
  Systems (TiiS)}} \bibinfo{volume}{3}, \bibinfo{number}{3}
  (\bibinfo{year}{2013}), \bibinfo{pages}{1--7}.
\newblock


\bibitem[\protect\citeauthoryear{Correa and Bareinboim}{Correa and
  Bareinboim}{2017}]%
        {correa2017causal}
\bibfield{author}{\bibinfo{person}{Juan~D Correa} {and} \bibinfo{person}{Elias
  Bareinboim}.} \bibinfo{year}{2017}\natexlab{}.
\newblock \showarticletitle{Causal effect identification by adjustment under
  confounding and selection biases}. In \bibinfo{booktitle}{\emph{Thirty-First
  AAAI Conference on Artificial Intelligence}}.
\newblock


\bibitem[\protect\citeauthoryear{Correa, Tian, and Bareinboim}{Correa
  et~al\mbox{.}}{2018}]%
        {correa-aaai18}
\bibfield{author}{\bibinfo{person}{Juan~D Correa}, \bibinfo{person}{Jin Tian},
  {and} \bibinfo{person}{Elias Bareinboim}.} \bibinfo{year}{2018}\natexlab{}.
\newblock \showarticletitle{Generalized adjustment under confounding and
  selection biases}. In \bibinfo{booktitle}{\emph{AAAI}}.
\newblock


\bibitem[\protect\citeauthoryear{Craswell, Zoeter, Taylor, and Ramsey}{Craswell
  et~al\mbox{.}}{2008}]%
        {craswell-wsdm08}
\bibfield{author}{\bibinfo{person}{Nick Craswell}, \bibinfo{person}{Onno
  Zoeter}, \bibinfo{person}{Michael Taylor}, {and} \bibinfo{person}{Bill
  Ramsey}.} \bibinfo{year}{2008}\natexlab{}.
\newblock \showarticletitle{An experimental comparison of click position-bias
  models}. In \bibinfo{booktitle}{\emph{WSDM}}. ACM, \bibinfo{pages}{87--94}.
\newblock


\bibitem[\protect\citeauthoryear{Dan-Dan, An, Ming-Sheng, and Jian}{Dan-Dan
  et~al\mbox{.}}{2013}]%
        {dan2013long}
\bibfield{author}{\bibinfo{person}{Zhao Dan-Dan}, \bibinfo{person}{Zeng An},
  \bibinfo{person}{Shang Ming-Sheng}, {and} \bibinfo{person}{Gao Jian}.}
  \bibinfo{year}{2013}\natexlab{}.
\newblock \showarticletitle{Long-term effects of recommendation on the
  evolution of online systems}.
\newblock \bibinfo{journal}{\emph{Chinese Physics Letters}}
  \bibinfo{volume}{30}, \bibinfo{number}{11} (\bibinfo{year}{2013}),
  \bibinfo{pages}{118901}.
\newblock


\bibitem[\protect\citeauthoryear{Dandekar, Goel, and Lee}{Dandekar
  et~al\mbox{.}}{2013}]%
        {dandekar2013biased}
\bibfield{author}{\bibinfo{person}{Pranav Dandekar}, \bibinfo{person}{Ashish
  Goel}, {and} \bibinfo{person}{David~T Lee}.} \bibinfo{year}{2013}\natexlab{}.
\newblock \showarticletitle{Biased assimilation, homophily, and the dynamics of
  polarization}.
\newblock \bibinfo{journal}{\emph{Proceedings of the National Academy of
  Sciences}} \bibinfo{volume}{110}, \bibinfo{number}{15}
  (\bibinfo{year}{2013}), \bibinfo{pages}{5791--5796}.
\newblock


\bibitem[\protect\citeauthoryear{Dwork, Kumar, Naor, and Sivakumar}{Dwork
  et~al\mbox{.}}{2001}]%
        {dwork-www01}
\bibfield{author}{\bibinfo{person}{Cynthia Dwork}, \bibinfo{person}{Ravi
  Kumar}, \bibinfo{person}{Moni Naor}, {and} \bibinfo{person}{Dandapani
  Sivakumar}.} \bibinfo{year}{2001}\natexlab{}.
\newblock \showarticletitle{Rank aggregation methods for the web}. In
  \bibinfo{booktitle}{\emph{Proceedings of the 10th international conference on
  World Wide Web}}. ACM, \bibinfo{pages}{613--622}.
\newblock


\bibitem[\protect\citeauthoryear{Fleder and Hosanagar}{Fleder and
  Hosanagar}{2007}]%
        {fleder2007recommender}
\bibfield{author}{\bibinfo{person}{Daniel~M Fleder} {and}
  \bibinfo{person}{Kartik Hosanagar}.} \bibinfo{year}{2007}\natexlab{}.
\newblock \showarticletitle{Recommender systems and their impact on sales
  diversity}. In \bibinfo{booktitle}{\emph{Proceedings of the 8th ACM
  conference on Electronic commerce}}. ACM, \bibinfo{pages}{192--199}.
\newblock


\bibitem[\protect\citeauthoryear{Heckman}{Heckman}{1979}]%
        {heckman-econometrica79}
\bibfield{author}{\bibinfo{person}{James Heckman}.}
  \bibinfo{year}{1979}\natexlab{}.
\newblock \showarticletitle{{Sample Selection Bias as a Specification Error}}.
\newblock \bibinfo{journal}{\emph{Econometrica}} \bibinfo{volume}{47},
  \bibinfo{number}{1} (\bibinfo{year}{1979}), \bibinfo{pages}{153--161}.
\newblock


\bibitem[\protect\citeauthoryear{Hern{\'a}ndez-Lobato, Houlsby, and
  Ghahramani}{Hern{\'a}ndez-Lobato et~al\mbox{.}}{2014}]%
        {hernandez2014probabilistic}
\bibfield{author}{\bibinfo{person}{Jos{\'e}~Miguel Hern{\'a}ndez-Lobato},
  \bibinfo{person}{Neil Houlsby}, {and} \bibinfo{person}{Zoubin Ghahramani}.}
  \bibinfo{year}{2014}\natexlab{}.
\newblock \showarticletitle{Probabilistic matrix factorization with non-random
  missing data}. In \bibinfo{booktitle}{\emph{International Conference on
  Machine Learning}}. \bibinfo{pages}{1512--1520}.
\newblock


\bibitem[\protect\citeauthoryear{Hofmann, Schuth, Whiteson, and
  de~Rijke}{Hofmann et~al\mbox{.}}{2013}]%
        {hofmann2013reusing}
\bibfield{author}{\bibinfo{person}{Katja Hofmann}, \bibinfo{person}{Anne
  Schuth}, \bibinfo{person}{Shimon Whiteson}, {and} \bibinfo{person}{Maarten de
  Rijke}.} \bibinfo{year}{2013}\natexlab{}.
\newblock \showarticletitle{Reusing historical interaction data for faster
  online learning to rank for IR}. In \bibinfo{booktitle}{\emph{Proceedings of
  the sixth ACM international conference on Web search and data mining}}. ACM,
  \bibinfo{pages}{183--192}.
\newblock


\bibitem[\protect\citeauthoryear{Hu, Wang, Peng, and Li}{Hu
  et~al\mbox{.}}{2019}]%
        {hu2019unbiased}
\bibfield{author}{\bibinfo{person}{Ziniu Hu}, \bibinfo{person}{Yang Wang},
  \bibinfo{person}{Qu Peng}, {and} \bibinfo{person}{Hang Li}.}
  \bibinfo{year}{2019}\natexlab{}.
\newblock \showarticletitle{Unbiased LambdaMART: An Unbiased Pairwise
  Learning-to-Rank Algorithm}.
\newblock  (\bibinfo{year}{2019}).
\newblock


\bibitem[\protect\citeauthoryear{Jagerman, Oosterhuis, and de~Rijke}{Jagerman
  et~al\mbox{.}}{2019}]%
        {jagerman2019model}
\bibfield{author}{\bibinfo{person}{Rolf Jagerman}, \bibinfo{person}{Harrie
  Oosterhuis}, {and} \bibinfo{person}{Maarten de Rijke}.}
  \bibinfo{year}{2019}\natexlab{}.
\newblock \showarticletitle{To Model or to Intervene: A Comparison of
  Counterfactual and Online Learning to Rank from User Interactions}.
\newblock  (\bibinfo{year}{2019}).
\newblock


\bibitem[\protect\citeauthoryear{Japec, Kreuter, Berg, Biemer, Decker, Lampe,
  Lane, O'Neil, and Usher}{Japec et~al\mbox{.}}{2015}]%
        {japec-poq15}
\bibfield{author}{\bibinfo{person}{Lilli Japec}, \bibinfo{person}{Frauke
  Kreuter}, \bibinfo{person}{Marcus Berg}, \bibinfo{person}{Paul Biemer},
  \bibinfo{person}{Paul Decker}, \bibinfo{person}{Cliff Lampe},
  \bibinfo{person}{Julia Lane}, \bibinfo{person}{Cathy O'Neil}, {and}
  \bibinfo{person}{Abe Usher}.} \bibinfo{year}{2015}\natexlab{}.
\newblock \showarticletitle{Big data in survey research: Aapor task force
  report}.
\newblock \bibinfo{journal}{\emph{Public Opinion Quarterly}}
  \bibinfo{volume}{79}, \bibinfo{number}{4} (\bibinfo{year}{2015}),
  \bibinfo{pages}{839--880}.
\newblock
\showISSN{0033-362X}
\urldef\tempurl%
\url{https://doi.org/10.1093/poq/nfv039}
\showDOI{\tempurl}


\bibitem[\protect\citeauthoryear{Joachims}{Joachims}{2002}]%
        {joachims2002optimizing}
\bibfield{author}{\bibinfo{person}{Thorsten Joachims}.}
  \bibinfo{year}{2002}\natexlab{}.
\newblock \showarticletitle{Optimizing search engines using clickthrough data}.
  In \bibinfo{booktitle}{\emph{Proceedings of the eighth ACM SIGKDD
  international conference on Knowledge discovery and data mining}}. ACM,
  \bibinfo{pages}{133--142}.
\newblock


\bibitem[\protect\citeauthoryear{Joachims, Granka, Pan, Hembrooke, and
  Gay}{Joachims et~al\mbox{.}}{2005}]%
        {joachims2005accurately}
\bibfield{author}{\bibinfo{person}{Thorsten Joachims}, \bibinfo{person}{Laura~A
  Granka}, \bibinfo{person}{Bing Pan}, \bibinfo{person}{Helene Hembrooke},
  {and} \bibinfo{person}{Geri Gay}.} \bibinfo{year}{2005}\natexlab{}.
\newblock \showarticletitle{Accurately interpreting clickthrough data as
  implicit feedback}. In \bibinfo{booktitle}{\emph{Sigir}},
  Vol.~\bibinfo{volume}{5}. \bibinfo{pages}{154--161}.
\newblock


\bibitem[\protect\citeauthoryear{Joachims, Swaminathan, and Schnabel}{Joachims
  et~al\mbox{.}}{2017}]%
        {joachims-wsdm17}
\bibfield{author}{\bibinfo{person}{Thorsten Joachims}, \bibinfo{person}{Adith
  Swaminathan}, {and} \bibinfo{person}{Tobias Schnabel}.}
  \bibinfo{year}{2017}\natexlab{}.
\newblock \showarticletitle{Unbiased learning-to-rank with biased feedback}. In
  \bibinfo{booktitle}{\emph{WSDM}}. ACM, \bibinfo{pages}{781--789}.
\newblock


\bibitem[\protect\citeauthoryear{Lazer, Kennedy, King, and Vespignani}{Lazer
  et~al\mbox{.}}{2014}]%
        {lazer-science14}
\bibfield{author}{\bibinfo{person}{David Lazer}, \bibinfo{person}{Ryan
  Kennedy}, \bibinfo{person}{Gary King}, {and} \bibinfo{person}{Alessandro
  Vespignani}.} \bibinfo{year}{2014}\natexlab{}.
\newblock \showarticletitle{The parable of Google Flu: traps in big data
  analysis}.
\newblock \bibinfo{journal}{\emph{Science}} \bibinfo{volume}{343},
  \bibinfo{number}{6176} (\bibinfo{year}{2014}), \bibinfo{pages}{1203--1205}.
\newblock


\bibitem[\protect\citeauthoryear{Lin}{Lin}{2010}]%
        {lin-wir2010}
\bibfield{author}{\bibinfo{person}{Shili Lin}.}
  \bibinfo{year}{2010}\natexlab{}.
\newblock \showarticletitle{Rank aggregation methods}.
\newblock \bibinfo{journal}{\emph{Wiley Interdisciplinary Reviews:
  Computational Statistics}} \bibinfo{volume}{2}, \bibinfo{number}{5}
  (\bibinfo{year}{2010}), \bibinfo{pages}{555--570}.
\newblock


\bibitem[\protect\citeauthoryear{Liu}{Liu}{2011}]%
        {liu-springer11}
\bibfield{author}{\bibinfo{person}{Tie-Yan Liu}.}
  \bibinfo{year}{2011}\natexlab{}.
\newblock \bibinfo{booktitle}{\emph{Learning to rank for information
  retrieval}}.
\newblock \bibinfo{publisher}{Springer Science \& Business Media}.
\newblock


\bibitem[\protect\citeauthoryear{Oosterhuis and de~Rijke}{Oosterhuis and
  de~Rijke}{2018}]%
        {oosterhuis2018differentiable}
\bibfield{author}{\bibinfo{person}{Harrie Oosterhuis} {and}
  \bibinfo{person}{Maarten de Rijke}.} \bibinfo{year}{2018}\natexlab{}.
\newblock \showarticletitle{Differentiable unbiased online learning to rank}.
  In \bibinfo{booktitle}{\emph{Proceedings of the 27th ACM International
  Conference on Information and Knowledge Management}}. ACM,
  \bibinfo{pages}{1293--1302}.
\newblock


\bibitem[\protect\citeauthoryear{Pearl, Glymour, and Jewell}{Pearl
  et~al\mbox{.}}{2016}]%
        {pearl2016causal}
\bibfield{author}{\bibinfo{person}{Judea Pearl}, \bibinfo{person}{Madelyn
  Glymour}, {and} \bibinfo{person}{Nicholas~P Jewell}.}
  \bibinfo{year}{2016}\natexlab{}.
\newblock \bibinfo{booktitle}{\emph{Causal inference in statistics: A primer}}.
\newblock \bibinfo{publisher}{John Wiley \& Sons}.
\newblock


\bibitem[\protect\citeauthoryear{Raman and Joachims}{Raman and
  Joachims}{2013}]%
        {raman2013learning}
\bibfield{author}{\bibinfo{person}{Karthik Raman} {and}
  \bibinfo{person}{Thorsten Joachims}.} \bibinfo{year}{2013}\natexlab{}.
\newblock \showarticletitle{Learning socially optimal information systems from
  egoistic users}. In \bibinfo{booktitle}{\emph{Joint European Conference on
  Machine Learning and Knowledge Discovery in Databases}}. Springer,
  \bibinfo{pages}{128--144}.
\newblock


\bibitem[\protect\citeauthoryear{Richardson, Dominowska, and Ragno}{Richardson
  et~al\mbox{.}}{2007}]%
        {richardson2007predicting}
\bibfield{author}{\bibinfo{person}{Matthew Richardson}, \bibinfo{person}{Ewa
  Dominowska}, {and} \bibinfo{person}{Robert Ragno}.}
  \bibinfo{year}{2007}\natexlab{}.
\newblock \showarticletitle{Predicting clicks: estimating the click-through
  rate for new ads}. In \bibinfo{booktitle}{\emph{WWW}}. ACM,
  \bibinfo{pages}{521--530}.
\newblock


\bibitem[\protect\citeauthoryear{Schnabel, Swaminathan, Singh, Chandak, and
  Joachims}{Schnabel et~al\mbox{.}}{2016}]%
        {schnabel2016recommendations}
\bibfield{author}{\bibinfo{person}{Tobias Schnabel}, \bibinfo{person}{Adith
  Swaminathan}, \bibinfo{person}{Ashudeep Singh}, \bibinfo{person}{Navin
  Chandak}, {and} \bibinfo{person}{Thorsten Joachims}.}
  \bibinfo{year}{2016}\natexlab{}.
\newblock \showarticletitle{Recommendations as treatments: Debiasing learning
  and evaluation}.
\newblock \bibinfo{journal}{\emph{arXiv preprint arXiv:1602.05352}}
  (\bibinfo{year}{2016}).
\newblock


\bibitem[\protect\citeauthoryear{Schuth, Oosterhuis, Whiteson, and
  de~Rijke}{Schuth et~al\mbox{.}}{2016}]%
        {schuth2016multileave}
\bibfield{author}{\bibinfo{person}{Anne Schuth}, \bibinfo{person}{Harrie
  Oosterhuis}, \bibinfo{person}{Shimon Whiteson}, {and}
  \bibinfo{person}{Maarten de Rijke}.} \bibinfo{year}{2016}\natexlab{}.
\newblock \showarticletitle{Multileave gradient descent for fast online
  learning to rank}. In \bibinfo{booktitle}{\emph{Proceedings of the Ninth ACM
  International Conference on Web Search and Data Mining}}. ACM,
  \bibinfo{pages}{457--466}.
\newblock


\bibitem[\protect\citeauthoryear{Schuth, Sietsma, Whiteson, Lefortier, and
  de~Rijke}{Schuth et~al\mbox{.}}{2014}]%
        {schuth2014multileaved}
\bibfield{author}{\bibinfo{person}{Anne Schuth}, \bibinfo{person}{Floor
  Sietsma}, \bibinfo{person}{Shimon Whiteson}, \bibinfo{person}{Damien
  Lefortier}, {and} \bibinfo{person}{Maarten de Rijke}.}
  \bibinfo{year}{2014}\natexlab{}.
\newblock \showarticletitle{Multileaved comparisons for fast online
  evaluation}. In \bibinfo{booktitle}{\emph{Proceedings of the 23rd ACM
  International Conference on Conference on Information and Knowledge
  Management}}. ACM, \bibinfo{pages}{71--80}.
\newblock


\bibitem[\protect\citeauthoryear{Smith and Elkan}{Smith and Elkan}{2004}]%
        {smith-kdd04}
\bibfield{author}{\bibinfo{person}{Andrew Smith} {and} \bibinfo{person}{Charles
  Elkan}.} \bibinfo{year}{2004}\natexlab{}.
\newblock \showarticletitle{A Bayesian network framework for reject inference}.
\newblock \bibinfo{journal}{\emph{KDD}} (\bibinfo{year}{2004}).
\newblock
\urldef\tempurl%
\url{http://delivery.acm.org/10.1145/1020000/1014085/p286-smith.pdf}
\showURL{%
\tempurl}


\bibitem[\protect\citeauthoryear{Wang, Bendersky, Metzler, and Najork}{Wang
  et~al\mbox{.}}{2016}]%
        {wang-sigir16}
\bibfield{author}{\bibinfo{person}{Xuanhui Wang}, \bibinfo{person}{Michael
  Bendersky}, \bibinfo{person}{Donald Metzler}, {and} \bibinfo{person}{Marc
  Najork}.} \bibinfo{year}{2016}\natexlab{}.
\newblock \showarticletitle{Learning to rank with selection bias in personal
  search}. In \bibinfo{booktitle}{\emph{SIGIR}}. ACM,
  \bibinfo{pages}{115--124}.
\newblock


\bibitem[\protect\citeauthoryear{Wang, Golbandi, Bendersky, Metzler, and
  Najork}{Wang et~al\mbox{.}}{2018a}]%
        {wang-wsdm18}
\bibfield{author}{\bibinfo{person}{Xuanhui Wang}, \bibinfo{person}{Nadav
  Golbandi}, \bibinfo{person}{Michael Bendersky}, \bibinfo{person}{Donald
  Metzler}, {and} \bibinfo{person}{Marc Najork}.}
  \bibinfo{year}{2018}\natexlab{a}.
\newblock \showarticletitle{Position bias estimation for unbiased learning to
  rank in personal search}. In \bibinfo{booktitle}{\emph{WSDM}}. ACM,
  \bibinfo{pages}{610--618}.
\newblock


\bibitem[\protect\citeauthoryear{Wang, Liang, Charlin, and Blei}{Wang
  et~al\mbox{.}}{2018b}]%
        {wang2018deconfounded}
\bibfield{author}{\bibinfo{person}{Yixin Wang}, \bibinfo{person}{Dawen Liang},
  \bibinfo{person}{Laurent Charlin}, {and} \bibinfo{person}{David~M Blei}.}
  \bibinfo{year}{2018}\natexlab{b}.
\newblock \showarticletitle{The deconfounded recommender: A causal inference
  approach to recommendation}.
\newblock \bibinfo{journal}{\emph{arXiv preprint arXiv:1808.06581}}
  (\bibinfo{year}{2018}).
\newblock


\bibitem[\protect\citeauthoryear{Yue and Joachims}{Yue and Joachims}{2009}]%
        {yue2009interactively}
\bibfield{author}{\bibinfo{person}{Yisong Yue} {and} \bibinfo{person}{Thorsten
  Joachims}.} \bibinfo{year}{2009}\natexlab{}.
\newblock \showarticletitle{Interactively optimizing information retrieval
  systems as a dueling bandits problem}. In
  \bibinfo{booktitle}{\emph{Proceedings of the 26th Annual International
  Conference on Machine Learning}}. ACM, \bibinfo{pages}{1201--1208}.
\newblock


\bibitem[\protect\citeauthoryear{Zadrozny}{Zadrozny}{2004}]%
        {zadrozny2004learning}
\bibfield{author}{\bibinfo{person}{Bianca Zadrozny}.}
  \bibinfo{year}{2004}\natexlab{}.
\newblock \showarticletitle{Learning and evaluating classifiers under sample
  selection bias}. In \bibinfo{booktitle}{\emph{Proceedings of the twenty-first
  international conference on Machine learning}}. ACM, \bibinfo{pages}{114}.
\newblock


\end{thebibliography}
\bibliographystyle{ACM-Reference-Format}

\end{document}